\documentclass[reprint,
superscriptaddress,
 amsmath,amssymb,
 aps]{revtex4-2}

\usepackage{graphicx}
\usepackage{dcolumn}
\usepackage{bm}
\usepackage{comment}

\usepackage[utf8]{inputenc}
\usepackage[T1]{fontenc}

\usepackage{etoolbox}
\usepackage[linesnumbered,ruled,vlined]{algorithm2e}
\usepackage{booktabs} 
\usepackage{overpic}
\usepackage{algorithmic}
\usepackage{amsmath}
\usepackage{color,soul}
\usepackage{xcolor,colortbl}
\usepackage{wrapfig}
\usepackage{amsfonts}
\usepackage{amsthm}

\usepackage{float}
\usepackage[section]{placeins}

\DeclareMathOperator*{\argmax}{argmax} 

\DeclareMathOperator{\argmin}{argmin}

\begin{document}

\preprint{AIP/123-QED}

\title{Model-free estimation in scattering analysis of microscopy}
\author{Tong Lin}
\affiliation{
Department of Statistics and Applied Probability, University of California, Santa Barbara, CA 93106, USA
}

\author{Jinseok Lee}
\affiliation{
Department of Mechanical Engineering, Yale University, New Haven, CT 06520, USA
}

\author{Matt Helgeson}
\affiliation{
Department of Chemical Engineering, University of California, Santa Barbara, CA 93106, USA
}

\author{Megan T. Valentine}
\affiliation{
Department of Mechanical Engineering, University of California, Santa Barbara, CA 93106, USA
}

\author{Yimin Luo}
\affiliation{
Department of Mechanical Engineering, Yale University, New Haven, CT 06520, USA
}

\author{Mengyang Gu}
\email{Corresponding author: mengyang@pstat.ucsb.edu}
\affiliation{
Department of Statistics and Applied Probability, University of California, Santa Barbara, CA 93106, USA
}

\date{\today}

\begin{abstract}

The mean squared displacement (MSD) of particles or probes is commonly estimated from microscopy videos using particle tracking approaches, which rely on tuning parameters manually, and are often unstable over the entire lag time range, especially in dense or low-contrast situations. 
In this work, we propose model-free \textit{ab initio} uncertainty quantification (MF-AIUQ), a model-free method for scattering analysis of microscopy video based on a probabilistic framework, which estimates MSD without isolating particles and linking their trajectories. Based on the relationship between the intermediate scattering function (ISF) and the MSD derived from the cumulant theorem, MF-AIUQ estimates the MSD values by the marginal maximum likelihood estimator. To reduce the computational cost, the likelihood function is approximated by a subset of Fourier-transformed intensities. These intensities are equally spaced at the logarithmic values of Fourier basis functions and lag time points. We found that the ISF is smooth in this logarithmic input space, and the information of the ISF can be captured by this subset of inputs. We examine the method through simulation studies covering several representative stochastic processes and three experimental systems: a Newtonian fluid for evaluating performance in optically dense and bright-field settings, a gelation system with an evolving MSD shape, and snail mucin, a viscoelastic biopolymer, for modulus estimation. Across these studies, MF-AIUQ provides smooth and stable MSD estimates over the full lag time range and serves as a useful complementary approach in settings where particle tracking is unreliable or a parametric model of MSD is unavailable or unverifiable.

\end{abstract}

\maketitle

\section{\label{sec:intro}INTRODUCTION}

Colloids undergo ceaseless motion as they are constantly bombarded by surrounding water molecules. This phenomenon drives processes such as the diffusion of nutrients, oxygen, and signaling molecules through cells and tissues. Mean squared displacement (MSD) is an important quantity for characterizing particle dynamics in microscopy experiments. It is widely used to quantify particle diffusion behavior and infer material properties such as viscoelasticity \cite{mason1995optical}. However, obtaining reliable MSD estimates of particles from microscopy videos remains challenging in practice as the estimation can be sensitive to measurement error \cite{michalet2010mean, berglund2010statistics},  limited observation time \cite{qian1991single}, and model specification \cite{gu2024ab}. These challenges motivate the development of robust frameworks for model-free estimation of MSD from microscopy videos.

Approaches for estimating MSD from microscopy videos can be broadly divided into real-space and reciprocal-space methods. In real-space analysis, particle dynamics are typically quantified by segmenting individual particles and linking their positions across time to reconstruct trajectories \cite{crocker1996methods}. This framework was originally developed in the context of single particle tracking, where individual trajectories are analyzed to characterize diffusion and transport behavior \cite{qian1991single, saxton1997single}, and was later extended to multiple particle tracking (MPT) \cite{mason1997particle}. MPT is a widely used microrheological method that provides ensemble-averaged measures of particle dynamics, including the MSD, and has been applied to various materials such as hydrogels \cite{luo2025optimizing}, colloidal gels \cite{zaccarelli2009colloidal,pickrahn2010relationship}, and stimuli-responsive materials \cite{mcglynn2020multiple}. MPT is highly effective when particles can be accurately identified and linked. However, its performance depends on user-specified parameters such as intensity thresholds, search radii, and maximum linking distances \cite{crocker1996methods,chenouard2014objective}, and can degrade in optically dense systems or low signal-to-noise conditions where reliable particle segmentation and trajectory reconstruction become increasingly difficult \cite{chenouard2014objective,savin2005static}. 

Reciprocal-space methods characterize particle dynamics by analyzing the temporal evolution of intensities in Fourier space. A representative approach is differential dynamic microscopy (DDM) \cite{cerbino2008differential}, which treats the microscope as a multi-angle scattering instrument and estimates particle dynamics by fitting a model to the image structure function, computed as the time-averaged, squared differences of Fourier-transformed images across wave vectors and lag times \cite{giavazzi2009scattering,lattuada2025hitchhiker}. By avoiding explicit particle segmentation and trajectory reconstruction, DDM serves as a useful alternative in optically dense or low-contrast settings where precise particle tracking methods prove difficult \cite{giavazzi2009scattering}. DDM has also been applied to diverse systems, including isolated or clustered proteins \cite{guidolin2023protein,safari2015differential}, active networks \cite{lee2021myosin}, colloidal gels \cite{giavazzi2016structure}, suspensions \cite{safari2017differential}, polymer hydrogels \cite{dhakal2026differential}, and motile bacteria \cite{wilson2011differential,lattuada2025hitchhiker,martinez2012differential}. Despite this tracking-free formulation, standard DDM analysis has two bottlenecks. First, manual selection of wave vector ranges and specific loss functions is often needed for parameter estimation in a case-by-case manner. Second, DDM often relies on a prespecified model. While model-free DDM analyses have been developed based on directly inverting the image structure function to obtain MSD estimation separately for each lag time point \cite{bayles2017probe}, this approach may only provide reliable MSD estimation for several lag time points for certain systems, due to the limited information of image pairs at long lag time points \cite{gu2021uncertainty}. 

To address these limitations, we introduce a model-free extension of the model-dependent \textit{ab initio} uncertainty quantification (MD-AIUQ) for scattering analysis of dynamics proposed in Ref. \cite{gu2024ab}, referred to as MF-AIUQ, for estimating MSD from microscopy image sequences over the entire lag time range. MD-AIUQ develops a probabilistic latent factor model of the original intensity in the real space, with factor loadings being the discrete Fourier basis functions, and the correlation of each latent factor process modeled by the intermediate scattering function of a prespecified model. The standard DDM estimation by fitting the image structure function was shown to be equivalent to minimizing the temporal autocorrelation in the Fourier space of this probabilistic model. A more efficient estimator, namely the maximum marginal likelihood estimation (MMLE), was proposed to estimate model parameters. Compared with standard DDM approaches, MD-AIUQ does not require directly computing the difference of image pairs at each lag time point; rather, the two-time correlation function in Fourier space is modeled by a likelihood function, which weighs the information naturally using different Fourier basis functions, and removes the need to specify a Fourier range for estimation required in DDM approaches \cite{gu2024ab}. Rather than estimating parameters of a prescribed stochastic process, MF-AIUQ represents the intermediate scattering function through its relation with the MSD derived from the cumulant theorem \cite{koppel1972analysis}, which does not require specifying a particular model of the underlying dynamics \textit{a priori}.

The model-free estimation problem is more challenging than the model-dependent estimation \cite{gu2024ab}, as the entire MSD curve needs to be estimated rather than only several parameters in a prespecified stochastic process. To improve estimation stability and accelerate computation, we optimize a subsampled set of MSD values at logarithmically spaced lag time points to reduce the dimension, and use Gaussian process regression to interpolate the MSD over the entire lag time range. Moreover, we also subsample wave vectors and use the generalized Schur algorithm for efficient covariance inversion and log determinant computation \cite{gohberg1972inversion, ammar1988superfast}, thereby substantially reducing the overall computational cost. The resulting MSD estimates are often smoother than MPT, reducing the need for ad hoc fitting in subsequent microrheological analysis through the generalized Stokes–Einstein relation.

The contribution of this work is robust likelihood-based framework of scattering analysis of microscopy videos to obtain model-free MSD estimation with uncertainty quantification. Compared with MPT, MF-AIUQ does not require particle trajectory reconstruction and the associated parameter tuning step. 
Compared with existing model-free DDM approaches \cite{bayles2017probe, gu2021uncertainty}, which depend on direct inversion of the image structure function separately at each wave vector, MF-AIUQ is capable of estimating MSD over all lag times jointly through borrowing information across Fourier basis functions and lag time points by the marginal likelihood function. Through a default subsampling strategy applied to all simulated and experimental scenarios, the computational cost of MF-AIUQ is comparable to MD-AIUQ and DDM-UQ approaches, shown in Fig. \ref{fig:computation_time}.  Extensive numerical studies of six simulation scenarios and three different experimental settings in Sec. \ref{sec:simulation} and Sec. \ref{sec:experiment} confirm the unique advantages of MF-AIUQ compared with the alternative approaches.  
Although MF-AIUQ is not intended to replace MPT, or MD-AIUQ, it serves as a complementary approach when parametric model assumptions are difficult to justify or particle tracking is unreliable. The new model-free estimation of MSD curves has now been integrated into the open-sourced AIUQ package \cite{AIUQ2026Rpackage}.

\section{\label{sec:background}BACKGROUND}
\subsection{Differential dynamic microscopy}

We begin by considering a microscopy video with
$n$ equally spaced time frames. Each frame contains $N=N_1\times N_2$ pixels of intensities represented by an $N_1\times N_2$ matrix $\mathbf Y(t)$, with $Y_{j_1,j_2}(t)$ representing the intensity at the 2D coordinate $\mathbf x_{j_1,j_2}$ for any $j_1=1,...,N_1$ and $j_2=1,...,N_2$. A typical microscopy video contains $N=500\times 500$ pixels and $n=500$ time points. We then normalize the pixel intensity to the range $[0,1]$.

Let $ \mathbf {\hat Y}(t)=\mathcal F[\mathbf Y(t)] $ denote the 2D discrete Fourier transform of each image computed by the fast Fourier transform (FFT) algorithm \cite{cooley1965algorithm}, where the $(k_1,k_2)$ term of $\mathbf{\hat Y}(t)$ is the complex-valued intensity at wave vector $\mathbf q_{j'_1,j'_2}=(q_{j'_1,1},q_{j'_2,2})$: 
\begin{align*}
\hat Y_{j'_1,j'_2}(t)=\sum^{N_1}_{j_1=1}\sum^{N_2}_{j_2=1}Y_{j_1,j_2}(t) e^{-2\pi i \left(\frac{(j_1-1)(j'_1-1)}{N_1}+\frac{(j_2-1)(j'_2-1)}{N_2}\right)}, 
\end{align*}
where $i$ denotes the complex number.

In differential dynamic microscopy (DDM) \cite{cerbino2008differential,giavazzi2009scattering}, the image structure function at any wave vector $\mathbf q$ and lag time $\Delta t$ is computed as the square of the absolute difference of Fourier transformed intensity at two time frames
\begin{align}
D(\mathbf q, \Delta t)&=\frac{1}{(n-n_{\Delta t})}\sum^{n-n_{\Delta t} }_{t=1}|\Delta \hat Y_\mathbf q(t, \Delta t)|^2,
\label{equ:image_sf}
\end{align}
where $\Delta \hat Y_{\mathbf q}(t, \Delta t)=\hat Y_{\mathbf q}(t+\Delta t)- \hat Y_{\mathbf q}(t)$ and $n_{\Delta t}=\Delta t/\Delta t_{min}$ with $\Delta t_{min}$ being the difference between two consecutive time frames.

For microscopy videos of an isotropic process, we denote an index set $\mathcal S_j=\{(j'_1,j'_2): q^2_{j'_1,1}+q^2_{j'_2,2}=q^2_j\}$ for $j=1,...,J$, which contains the indices of the $j$th `ring' of the Fourier-transformed quantity with amplitude $q_j=\frac{2\pi j}{\Delta x_{min} \sqrt{N}}$, with $\Delta x_{min}$ being the length of a pixel in one coordinate. For a microscopy image with a square field of view, $J=\lceil \sqrt{N}/2\rceil-1$ with $\lceil \sqrt{N}/2 \rceil$ being the smallest integer larger than or equal to  $\sqrt{N}/2$. 
Denoting $N_{S_j}$ the number of elements in the $j$th ring index set $\mathcal S_j$.  The average image structure function at the $j$th ring follows 
\begin{equation}
D(q_j, \Delta t)=\frac{1}{N_{S_j}}\sum_{\mathbf j' \in \mathcal S_j} D(\mathbf q_{\mathbf j'}, \Delta t). 
\label{equ:image_sf_compute}
\end{equation}
The model of this quantity follows decomposition 
\begin{align}
D_m(q_j, \Delta t)&=A_j (1-f_{\bm \theta}(q_j, \Delta t))+ B,
\label{equ:image_sf_model}
\end{align}
where $f_{\bm \theta}(q_j, \Delta t)$ is the intermediate scattering function (ISF) at ring $j$ and parameters $\bm \theta$. Parameters $A_j$ and $B$ represent amplitude and twice the variance of the background noise, respectively, which are often estimated from the data. The ISF is a crucial quantity that characterizes particles' self-evolution over time, while only a limited number of processes have closed-form expressions of the ISF. For instance, for a 2D isotropic Brownian motion or diffusion, the ISF follows $f(q,\Delta t)=\exp(-q^2 \mathcal{D} \Delta t)$ where $\mathcal{D}$ is the diffusion coefficient. 

In many complex systems, however, the ISF is unknown. By applying the cumulant theorem, the ISF can be approximated by a function of MSD at any lag time $\Delta t$ \cite{ koppel1972analysis, nijboer1966time}:
\begin{equation}
    f_{\bm \theta}(\mathbf q,\Delta t) \approx
\exp\!\left(-\frac{q^2\,\theta_{\Delta t}}{4}\right),
\label{equ:isf_msd}
\end{equation}
where $\theta_{\Delta t}=\langle \Delta x^2(\Delta t)\rangle$ denotes the MSD value as a function $\Delta t$, with the subscript being the corresponding lag time. The derivation of (\ref{equ:isf_msd}) can be found in Appendix A in Ref. \cite{gu2024ab}. Now, as Eq. (\ref{equ:isf_msd}) does not rely on specifying a certain model, the estimation of ISF can be made in a model-free manner.  

Let $\bm{\theta}$ denote the collection of parameters that determine the ISF. In DDM, a subset of Fourier basis functions, denoted as $\mathcal J_s \subset \{1,...,J\}$, is often selected, where
$N_{J_s}$ denotes the number of selected rings. The parameters are estimated by minimizing a loss function between the observed and modeled image structure functions in Eqs. (\ref{equ:image_sf_compute})-(\ref{equ:image_sf_model}): 
 \begin{align}
&(\bm \theta_{est}, \mathbf A_{est,\mathcal J_s},  B_{est})=\underset{\bm \theta, \mathbf A_{\mathcal J_s},  B}{\argmin} \, \mbox{Loss}( \mathbf D, \mathbf D_m), 
\label{equ:loss_Dqt}
\end{align}
where $\mathbf D$ and $\mathbf D_m$ are $N_{J_s}\times (n-1)$ matrices of observed and modeled image structure functions, with the $(j,k)$th term being $D(q_j,\Delta t_{k})$ and  $D_m(q_j,\Delta t_{k})$  respectively, for $j\in \mathcal J_s$ and $k=1,...,n-1$. A frequently used loss function is the $L_2$ loss, given by $\mbox{Loss}( \mathbf D, \mathbf D_m)=||\mathbf D-  \mathbf D_m||^2$ where $||\cdot ||$ denotes the $L_2$ distance.

Still, parameter estimation in Eq.  (\ref{equ:loss_Dqt}) can be challenging due to the need to pre-specify a Fourier range or a parametric model of ISF. A small number of studies in DDM allow for model-free estimation in some systems \cite{bayles2017probe,gu2021uncertainty}, based on Eq. (\ref{equ:isf_msd}). 
One such model-free DDM approach is differential dynamic microscopy with uncertainty quantification (DDM-UQ) \cite{gu2021uncertainty}, which estimates the MSD through the direct inversion (DI)  of image structure function \cite{bayles2017probe}. In DDM-UQ, the noise parameter $B$ is estimated by
\begin{equation}
  B^{DI}=
\min\Bigl\{
\left\langle D(q_{J},\Delta t)\right\rangle_{\Delta t},\, \min_{j \in 1:J} D(q_j,\Delta t_{1})
\Bigr\},
\label{equ:B_DI}  
\end{equation}
where $\left\langle D(q_{J},\Delta t)\right\rangle_{\Delta t}$ represents the average image structure function over lag times at the largest wave vector, and $\min_{j \in 1:J} D(q_j,\Delta t_{1})$ is the minimum image structure function across the wave vectors at the smallest lag time. Given $B^{DI}$, the amplitude parameters are estimated using the unbiased estimator defined later in Eq. (\ref{equ:A_est}), denoted by $\mathbf{A}^{DI}_{1:J}$. The corresponding MSD estimates for each wave vector and lag time are obtained using Eqs. (\ref{equ:image_sf_model}) and (\ref{equ:isf_msd}). Denote the resulting matrix of MSD estimates by $\bm \Theta^{DI}$. For $j \in \mathcal{J}_{s}$, where $\mathcal{J}_{s}$ represents a selected wave vector range, the $(j,k)$th entry of $\bm \Theta^{DI}$ is 
\begin{align}
\Theta^{DI}(q_j, \Delta t_k) 
& =  \log\!\left(
\frac{A^{DI}_{j}}
{ A^{DI}_{j}-D^{DI}(q_j,\Delta t_k)+B^{DI}}\right) \nonumber\\
& \quad \times \frac{4}{q_{j}^2}.
\label{equ:msd_DI}
\end{align}
The medians across wave vectors are used to estimate MSD at different lag times. However, the DDM-UQ approach can only reliably estimate MSD information for a short range of lag time points. To overcome the limit of unstable model fitting, a recent probabilistic framework was proposed in Ref. \cite{gu2024ab}, which 
naturally weights information at different lag time points and Fourier basis functions through the  likelihood function, yet it requires specifying a parametric model of the ISF. We briefly review this probabilistic framework in Section \ref{subsec:latent_model}.

\subsection{Probabilistic representation of microscopy videos with spatial Fourier basis}
\label{subsec:latent_model}

Let $\mathbf y(t)=\mbox{Vec}[\mathbf Y(t)]$ denote the $N$-dimensional vector of a microscopy image at any time frame $t$ and $\mbox{Vec}[\cdot]$ denotes the vectorization operation. As shown in Ref. \cite{gu2024ab}, the image intensity is modeled by a latent factor model,
\begin{equation}
\mathbf{y}(t)
= \frac{1}{\sqrt{N}} \mathbf{W}^* \mathbf{z}(t)
+ \bm{\epsilon}(t),
\label{equ:latent_factor_model}
\end{equation}
where $\mathbf{W}^*$ is the $N\times N$ complex conjugate of the two-dimensional discrete Fourier basis matrix, $\mathbf{z}(t)$ is an $N$-dimensional complex-valued latent factor in reciprocal space, and $\bm{\epsilon}(t) \sim \mathcal{MN}(\mathbf{0}, \frac{{B}}{2}\mathbf{I}_N)$ is a Gaussian noise vector with variance $\frac{B}{2}$ and $\mathbf I_N$ denotes the identity matrix with $N$ dimensions.

 The latent factor model can be split into the vectors of real and imaginary components $\mathbf{z}(t) = \mathbf{z}_{re}(t) + i\,\mathbf{z}_{im}(t)$. Consider a microscopy video of $n$ time points at $\{t_{1},...,t_n\}$. For each wave vector ring $j=1,\ldots,J$, and  $\mathbf j' \in \mathcal S_j$, the transpose of the corresponding rows  in real and complex latent factor matrices $[\mathbf{z}_{re}(t_1),...,\mathbf{z}_{re}(t_n)]$ and $[\mathbf{z}_{im}(t_1),...,\mathbf{z}_{im}(t_n)]$  are assumed to follow a multivariate normal distribution. 
\[
\mathbf{z}_{re,\, \mathbf j'} \sim \mathcal{MN}\!\left(\mathbf{0}, \frac{A_j}{4}\mathbf{R}_j\right) 
\, \mbox{and} \, 
\mathbf{z}_{im,\, \mathbf j'} \sim \mathcal{MN}\!\left(\mathbf{0}, \frac{A_j}{4}\mathbf{R}_j\right),
\]
where $A_j$ is an amplitude parameter, and $\mathbf{R}_j$ is the $n \times n$ temporal correlation matrix encoding the two-time correlation function in the Fourier space \cite{brown1997speckle}.  
The $(k_1,k_2)$th entry of $\mathbf R_j$ is determined by the ISF:
\begin{equation}
R_j(k_1,k_2)
= f_{\bm \theta}(q_j,\Delta t_k),
\label{equ:corr}
\end{equation}
with $\Delta t_k = |k_2-k_1|\Delta t_{min}$ for any $k_1=1,...,n$ and $k_2=1,...,n$.

To connect this latent factor formulation with physically interpretable quantities used in DDM, one can multiply both sides of Eq.~\eqref{equ:latent_factor_model} by $\mathbf W/\sqrt{N}$ 
\begin{equation}
\hat{\mathbf y}(t)=\mathbf W \mathbf y(t)/\sqrt{N}.
\label{equ:transformed_y}
\end{equation}
For a given wave vector $\mathbf q$ and lag time $\Delta t$, the difference of the Fourier intensities over $\Delta t$ is  $\Delta \hat y_{\mathbf q}(t,\Delta t)
=\hat y_{\mathbf q}(t+\Delta t)-\hat y_{\mathbf q}(t)$. In Ref. \cite{gu2024ab}, the expectation of the squared difference of the Fourier-transformed intensity between any two time frames is shown to follow
\begin{align*}
&\mathbb E\left[
\left(\hat y_{\mathbf q}(t+\Delta t)-\hat y_{\mathbf q}(t)\right)
\left(\hat y_{\mathbf q}^{*}(t+\Delta t)-\hat y_{\mathbf q}^{*}(t)\right)
\right] \\
 = &A(\mathbf q)\bigl(1-f_{\bm \theta}(\mathbf q,\Delta t)\bigr)+ B \\
= &D_m(\mathbf q,\Delta t),
\label{equ:connection_ddm_aiqu}
\end{align*}
which is the modeled image structure function in Eq.~(\ref{equ:image_sf_model}) used for model fitting in DDM.  As model fitting in DDM estimation based on Eq. (\ref{equ:loss_Dqt}) is equivalent to minimizing the temporal variogram of the Fourier transformed image stack, Eq.  (\ref{equ:latent_factor_model}) provides a probabilistic representation of microscopy videos, enabling the development of efficient estimators and uncertainty propagation. 

Given a particular physical model, such as the Brownian motion (BM) or Ornstein-Uhlenbeck (OU) process, the MMLE of the parameters in the ISF after integrating out latent factor was studied in MD-AIUQ \cite{gu2024ab}, as the probabilistic model for microscopy videos is defined for the original observations at the beginning of the data processing. The goal of this work is to extend this probabilistic representation to model-free estimation, with the MSD itself directly being estimated from the microscopy video, which is more challenging as the number of parameters in model-free estimation is much larger.

\section{MODEL-FREE ESTIMATION}
\label{sec:methodology}
Here we develop model-free estimation from the AIUQ framework, referred as the MF-AIUQ,  based on Eq. (\ref{equ:latent_factor_model}), where the parameters in ISF is the $n-1$ MSD vector over the entire lag time range: $\bm \theta=(\theta_{\Delta t_1},...,\theta_{\Delta t_{n-1}})^T$. Denote $\hat{\mathbf y}_{re,\,\mathbf j'}$ and  $\hat{\mathbf y}_{im,\,\mathbf j'}$ as the real and imaginary vectors of the transpose of a row of Fourier transformed intensity corresponding to a pixel with index $\mathbf j' \in \mathcal S_j$ in the matrix $\{\hat{\mathbf y}(t_1),...,\hat{\mathbf y}(t_n)\}$ defined by Eq. (\ref{equ:transformed_y}). 
We follow Ref. \cite{gu2024ab} to integrate out the random latent factors $\{\mathbf z(t_1),...,\mathbf z(t_n)\}$, where the marginal likelihood of the parameters can be written as 
\begin{equation}
\begin{aligned}
\mathcal{L}(\bm{\theta},\mathbf{A}_{1:J},{B})
&=
\prod_{j=1}^{J}
\prod_{\mathbf j'\in\mathcal{S}_j}
p_{{MN}}\!\left(
\hat{\mathbf{y}}_{re,\,\mathbf  j'};\mathbf{0}_n,\bm{\Sigma}_j
\right) \\
&\quad\times
p_{{MN}}\!\left(
\hat{\mathbf{y}}_{im,\,\mathbf  j'};\mathbf{0}_n,\bm{\Sigma}_j
\right).
\end{aligned}
\label{eq:likelihood}
\end{equation}
where $p_{{MN}}(\mathbf{y}; \mathbf 0_n, \bm \Sigma_j)$ denotes the density of a multivariate normal distribution at any real-valued vector $\mathbf y$ with $n$-dimensional mean vector $\mathbf{0}_n$ and $n\times n$ covariance matrix 
\begin{equation}
    \bm{\Sigma}_j = \frac{A_j}{4}\mathbf{R}_j + \frac{{B}}{4}\mathbf{I}_n.
    \label{equ:covariance_mat}
\end{equation}

As the microscopy video is typically equally spaced in time, the matrix $\mathbf R_j$ is a Toeplitz matrix for all $j=1,...,J$. The generalized Schur algorithm \cite{gohberg1972inversion,ammar1988superfast, ling2019superfast} will be used to reduce the computational complexity of the log density of the multivariate normal distribution $\log(p_{MN}({\mathbf{y}}; \mathbf 0_n, \bm \Sigma_j))$ from $\mathcal O(n^3)$ operations to $\mathcal O(n\log^2(n))$ operations without approximation. 

Given any parameter $B$, an unbiased estimator of $A_j$ follows \cite{gu2024ab}
\begin{equation}
A_{est,j} =
\frac{2}{N_{S_j} n}
\sum_{\mathbf j'\in\mathcal{S}_j}
\sum_{k=1}^{n}
\bigl|\hat{y}_{\mathbf q_{\mathbf j'}}(t_k)\bigr|^2
- {B},
\label{equ:A_est}
\end{equation}

Unlike the model-dependent scenario in Ref. \cite{gu2024ab}, where the parameter vector $\bm \theta$ usually has only 1-3 dimensions, the vector $\bm \theta$ in model-free estimation represents the MSD curve at all lag times points and is $(n-1)$-dimensional. Directly optimizing $(\bm \theta,B)$ is computationally demanding and often numerically unstable. To address these challenges, we develop a new data reduction strategy with Gaussian process interpolation to reliably estimate the MSD at all lag time points. Compared with the DDM-UQ approach \cite{gu2021uncertainty}, this strategy enlarges the lag time range where MSD estimation can be obtained and improves precision, while also extending the AIUQ estimation in Ref. \cite{gu2024ab} to model-free scenarios.

\subsection{Data reduction through subsampling}

\begin{figure*}[!ht]
    \centering
    \begin{tabular}{cc}
     \multicolumn{2}{c}{
        \begin{overpic}[scale=0.8]{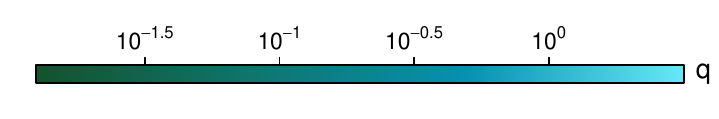}
        \end{overpic}
    } \\[-1em]
        \begin{overpic}[scale=0.68]{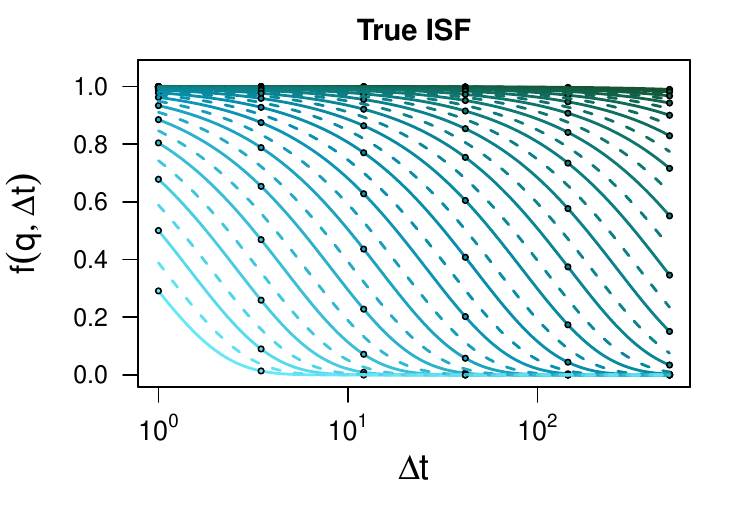}
            \put(0.5,65){\textbf{(a)}}
        \end{overpic} &
        \begin{overpic}[scale=0.68]{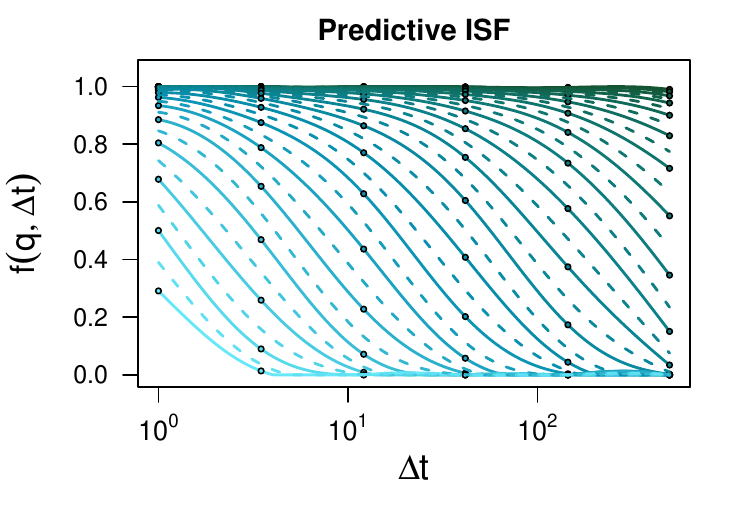}
            \put(0.5,65){\textbf{(b)}}
        \end{overpic} 
    \end{tabular}
    \caption{True and predicted intermediate scattering function curves for simulated BM with $\sigma_{BM}^2 = 0.5$. The circles represent the true ISF values used for training the GP model, corresponding to the 20 subsampled wave vectors and 6 subsampled lag times. Solid lines represent ISF curves at the subsampled $q$ values, while dashed lines represent ISF curves at several intermediate $q$ values not included in the training set.}
    \label{fig:ISF_plot}
\end{figure*}

To reduce computational complexity while preserving essential dynamical information, we develop a principled way to subsample the data in both the spatial and temporal domains.

In Ref. \cite{gu2024ab}, the default approach is to use all $J$ rings in the MMLE. However, even with acceleration by the generalized Schur algorithm, a single likelihood evaluation over all rings requires $\mathcal O(\sum^J_{j=1} N_{S_j}n \log^2(n))$ operations, where $N_{S_j}$ is the number of pixels in the $j$th ring of the Fourier transformed image and $n$ is the number of time points. Since likelihood optimization requires repeated evaluations of this quantity, the computational cost becomes substantial.

Here we first select an initial set of rings by choosing the smallest $J_0$ rings such that the cumulative proportion of amplitude reaches a threshold. This criterion is conventionally used to choose the number of  components in principal component analysis \cite{jolliffe2002principal} and has been used in the model-dependent AIUQ approach \cite{gu2024ab} 
\begin{equation*}
   \frac{\sum_{j=1}^{J_0} A_j}{\sum_{j=1}^{J} A_j} \ge 1-\varepsilon_1,
\end{equation*}
where $\varepsilon_1$ controls the proportion of total variation to be included. This criterion retains the rings that explain most of the signal variation while removing high-frequency components that are primarily dominated by complicated background noise, such as imaging artifacts (camera noise, out of focus light, etc). However, in systems with very slow particle dynamics, an informative signal may still appear in higher-frequency rings, where the amplitude increases again. To avoid discarding such rings, we further examine the tail rings with indices $J_0+1,...,J$ and update $J_0$ to the largest index satisfying $A_{J_0+1:J} \ge \varepsilon_2$. For most videos analyzed in this work, this step does not change the selection of $J_0$. For all numerical analyses used in this article, we let $\varepsilon_1=\varepsilon_2=0.001$.

From the $J_0$ pre-selected rings, we further subsample at most $N_s$ representative spatial frequencies with indices approximately equally spaced on the logarithmic scale, with spacing
\begin{equation}
\Delta_{\log q} = \frac{ \log(a J_0) }{N_s-1}.
\label{equ:delta_log_q}
\end{equation}
Then $\mathcal{J}_s$ represents the unique indices in the index set  $\{ 1, \lceil\exp(\Delta_{\log q })\rceil,..., \lceil\exp((N_s-1)\Delta_{\log q })\rceil \}$, where  $\lceil x\rceil$ denotes the smallest integer greater than or equal to $x$ and $a$ is slightly smaller than 1, which avoids the boundary effects from FFT if all rings are used. This strategy retains more rings at lower spatial frequencies, which often carry stronger and more informative dynamical signals, while sparsely sampling higher frequencies with limited variation.  We let $N_s=20$ and $c=0.95$ in all numerical studies. 

Similarly, we subsample the time domain through a coarse-graining of the MSD over lag times, which substantially reduces the optimization cost for jointly estimating the high-dimensional MSD parameters. We follow a strategy analogous to the spatial frequency subsampling and choose lag times that are approximately equally spaced on the logarithmic scale, with at most $n_{s}$ lag time points. The spacing is 
\begin{equation}
\Delta_{\log \Delta t} = \frac{ \log(n-1) }{n_s-1}.
\label{equ:delta_log_dt}
\end{equation}
We then define $\mathcal{T}_s$ as the set of unique indices in $\{ 1, \lceil\exp(\Delta_{\log \Delta t })\rceil,..., \lceil\exp((n_s-1)\Delta_{\log \Delta t  })\rceil \}$. The resulting MSD parameters are denoted as $\bm \theta_{\mathcal T_s}= (\theta_{\Delta  \tilde t_1},...,\theta_{\Delta  \tilde t_{\tilde n_s}})^T$ where 
$\{\Delta {\tilde t_1},...,\Delta  \tilde t_{\tilde n_s}\}=\Delta t_{min} \mathcal{T}_s$ are the $\tilde n_s$ unique lag time points selected. For all numerical studies, we let $n_s=\tilde n_s=6$. 
Then we interpolate the MSD over the entire range through the robust Gaussian process regression (GPR) \cite{Gu2018robustness}, based on robust marginal posterior mode estimation \cite{gu2018jointly}, by using the RobustGaSP package \cite{gu2018robustgasp} for computing the likelihood in Eq. (\ref{eq:likelihood}).

We found that the information on the reduced grid of wave vectors and lag times can capture the  ISF remarkably well. 
Figure \ref{fig:ISF_plot} compares the true and predicted ISF curves for a simulated BM with $\sigma_{BM}^2 = 0.5$, with simulation details discussed in Appendix \ref{app:simulation}. Since $f_{\bm \theta}(q,\Delta t)$ varies smoothly over $\log q$ and $\log \Delta t$, we apply a GPR model with input pairs $(\log q,\,  \log \Delta t)$ and response $f_{\bm \theta}(q,\Delta t)$. The training set consists of a reduced grid with $N_s = 20$ subsampled wave vectors and $n_s = 6$ subsampled lag times, shown as the circles in Fig. \ref{fig:ISF_plot}. The ISF surface is then predicted over the full logarithmic grid of wave vectors and lag times. 
In Fig. \ref{fig:ISF_plot}, the solid curves that correspond to ISF curves at the subsampled $q$, while dashed curves represent ISF curves at intermediate $q$ values not included in the training set. 
The close alignment between the predicted and true curves suggests that the selected sparse grid captures the main spatiotemporal structure of the ISF.

\subsection{Initialization and coarse-to-fine optimization}
\label{subset:ini_optim}
The parameters that require numerical optimization include the MSD at time index set $\mathcal T_s$, denoted as $\bm \theta_{\mathcal T_s}$, and the noise parameter $B$, where $\tilde{\bm\theta}_{\mathcal T_s}= [\log(\theta_{\Delta \tilde t_1}),...,\log(\theta_{\Delta   \tilde t_{\tilde n_s}})]^T$ is a vector of log MSD at $\tilde n_s$ time points and $\tilde B=\log(B/2)$ is a log variance of the noise of the intensity.

To improve numerical stability, we initialize the optimization using preliminary parameter estimates derived from 
Eq. (\ref{equ:msd_DI}) at the first two lag time points, denoted by $\log(\theta^{DI}_{\Delta t_1})$ and $\log(\theta^{DI}_{\Delta t_2})$. Computing these two quantities only requires calculating the image structure functions at the first two lag times, which is fast in practice, and the estimates at the two smallest time points are often stable \cite{gu2021uncertainty}. Given these two estimates, the initial values for $\tilde{\bm \theta}_{\mathcal{T}_s}$ are constructed by linearly interpolating a power-law trend on the logarithmic scale, 
\begin{equation}
\begin{aligned}
\tilde{\theta}^{(0)}_{\Delta  \tilde{t}_l}
& = \log \theta^{DI}_{\Delta t_1} + b\times \,
\frac{\log \theta^{DI}_{\Delta t_2} - \log \theta^{DI}_{\Delta t_1}}{\log \Delta t_2 - \log \Delta t_1} \\
&\qquad \times \bigl(\log \Delta \tilde{t}_l - \log \Delta t_1 \bigr),
\label{equ:ini_tilde_theta}
\end{aligned}
\end{equation}
where $l = 1, \dots, \tilde{n}_s$ and $b=0.9$ is used for all numerical studies.

For the noise parameter $\tilde{B}$, we initialize its value as the minimum over all Fourier rings of the logarithm of the spatially and temporally averaged squared Fourier intensities:
\begin{equation}
\tilde B^{(0)} = \min_{j=1,\ldots,J} \log \Bigl\{\frac{1}{N_{S_j} n} \sum_{\mathbf j' \in \mathcal{S}_j} \sum_{k=1}^{n} | \hat{y}_{\mathbf q_{\mathbf{j}'}}(t_k)|^2 \Bigr\}.
\label{equ:B_ini}
\end{equation}

We apply a coarse-to-fine optimization strategy for two time sequences with different resolutions. In the first stage, likelihood over a reduced lag time range is optimized to lower the computational cost and obtain a coarse parameter estimate. In the second stage, this estimate is refined by optimizing the likelihood over the entire temporal range. The L-BFGS algorithm \cite{liu1989limited} is used to both stages of the optimization.

{{The first-stage optimization is initialized at $\{\tilde{\bm\theta}^{(0)}_{\mathcal T_s},\, \tilde{B}^{(0)}\}$. At each iteration of the optimization, we run a GPR  trained with inputs $\log (\Delta t_{\min} \mathcal T_s) $ and outputs $\tilde{\bm\theta}_{\mathcal T_s}$, and predict over a coarse lag time grid in the logarithmic space $\{0, \log(c\Delta t_{\min}),..., \log(\lfloor(n-1)/c \rfloor c\Delta t_{\min}) \}$, with $\lfloor x \rfloor$ being the largest integer smaller than or equal to $x$.}} The value of $c$ is chosen adaptively according to the lag time length, and is set to $c = \lfloor n/100 \rfloor$ in all numerical studies with $n$ being the total number of time points. The set of lag time indices in this grid is defined as $\mathcal T_c=\{0,c,2c,...,\lfloor(n-1)/c \rfloor c \}$, where the first index corresponds to $\Delta t=0$.

The interpolated values are used to calculate a reduced likelihood, denoted by $\mathcal{L}_{\mathcal{T}_c,\, \mathcal{J}_s}$, which is evaluated over the spatial frequencies indexed by $\mathcal{J}_s$ and the equally spaced lag times indexed by $\mathcal{T}_c$. This retains the overall temporal trend of the MSD while significantly reducing the computational cost. The  estimate from the coarse-grained time points is obtained by
\[
(\tilde{\bm \theta}^{c}_{est,\mathcal{T}_s},\, \tilde{B}^{c}_{est})
=
\argmax_{\tilde{\bm \theta}_{\mathcal{T}_s},\, \tilde{B}}
\mathcal{L}_{\mathcal T_c,\,\mathcal J_s}
\!\left(\tilde{\bm \theta}_{\mathcal{T}_s},
\mathbf{A}_{est,\mathcal J_s}, \tilde{B}\right),
\]
where  given any $\tilde B$, the vector of amplitude parameters $\mathbf{A}_{est,\mathcal{J}_s}$ is obtained using the unbiased estimator in Eq. (\ref{equ:A_est}). 

The second stage is initialized at the estimates $\{\tilde{\bm \theta}^{c}_{est,\mathcal{T}_s},\, \tilde{B}^{c}_{est}\}$ obtained from stage one. GPR is applied in each iteration to interpolate the log MSD values over the full set of lag times indexed by $\mathcal{T}$, where $\mathcal{T} = \{0, 1,\dots, n-1 \}$ and the first index represents $\Delta t=0$. The likelihood evaluated over $\mathcal{J}_s$ and $\mathcal {T}$ is optimized to obtain the final estimate of the parameters
\[
(\tilde{\bm \theta}_{est,\mathcal{T}_s},\, \tilde{B}_{est})
=
\argmax_{\tilde{\bm \theta}_{\mathcal{T}_s},\, \tilde{B}}
\mathcal{L}_{\mathcal{T},\,\mathcal{J}_s}
\!\left(\tilde{\bm \theta}_{\mathcal{T}_s},
\mathbf{A}_{est, \mathcal{J}_s}, \tilde{B}\right).
\]
Through  the prediction of the GPR and exponential transformation, 
we obtain the MSD curve, denoted as $\bm{\theta}_{est} = (\theta_{est,\Delta t_1}, \dots, \theta_{est,\Delta t_{n-1}})^T$ at the entire lag time range. We consider two sources of estimation uncertainty, including Fourier discretization and parameter estimation error, with the details provided in Appendix \ref{app:MSD_uncertainty}. An overall algorithm for MF-AIUQ estimation is summarized in Algorithm \ref{alg:MF_AIUQ} in  Appendix \ref{app:MSD_uncertainty}.

\section{ESTIMATION FOR STORAGE AND LOSS MODULI}
\label{sec:storage_loss_modulus}
In cases where the particle dynamics are reporting the continuum properties of a homogeneous viscoelastic material, the MSD estimates can be further applied to microrheological analysis \cite{mason1995optical}. Specifically, the frequency-dependent storage and loss moduli, which characterize the viscoelastic responses of the material, are related to the MSD through the generalized Stokes-Einstein relation (GSER) and can thus be directly obtained \cite{mason2000estimating}. Earlier microrheological analyses estimated viscoelastic moduli by transforming the MSD into Laplace space, fitting an assumed functional form, and numerically converting the result into Fourier space for comparison with bulk rheology. However, this approach is sensitive to numerical truncation errors and functional assumptions for fitting MSD \cite{dasgupta2002microrheology, mason1995optical}. Later, a local power-law approach was proposed that avoids numerical transforms by estimating the complex modulus directly from the logarithmic time derivative of the MSD \cite{mason2000estimating}. 

Following the power-law approximation used in microrheology \cite{mason2000estimating, dasgupta2002microrheology}, the MSD estimate at lag time $\Delta t_j$ can be approximated by ${\theta}_{\Delta t_j}\,\propto\,(\Delta t_j)^{{\alpha}_{\Delta t_j}}$, with $j = 1,\dots, n-1$. Let $\tilde{\theta}_{\Delta t} = \log \theta_{\Delta t}$, where $\Delta t$ represents any lag time, and $\tilde{\bm \theta}_{est} = [\tilde{\theta}_{est, \Delta t_1}, \dots, \tilde{\theta}_{est, \Delta t_{n-1}}]^T = [\log {\theta}_{est, \Delta t_1}, \dots, \log {\theta}_{est, \Delta t_{n-1}}]^T$. The logarithmic derivative describes the slope of the MSD curve on a log–log scale, following
\[
{\alpha}_{\Delta t_j} \,=\, \frac{d\,\log{\theta}_{\Delta t_j}}{d\,\log\Delta t_j} \,= \, \frac{d\,\tilde{\theta}_{\Delta t_j}}{d\,\log\Delta t_j},
\]
where $d\, \cdot$ denotes the differential operator. 

For a 2D space, $j = 1,\dots,n-1$, the complex modulus is defined as
\begin{equation}
|G^{*}_{\omega_j}|
= \frac{2 k_B T_a}{3 \pi r \, {\theta}_{\Delta t_j}\,\Gamma\,\bigl(1+{\alpha}_{\Delta t_j}\bigr)},
\label{equ:G_complex}
\end{equation}
where $k_B$ is the Boltzmann constant, $T_a$ is the absolute temperature, $r$ is the particle radius, $\omega_j = 1/\Delta t_j$ is the frequency, and $\Gamma(\cdot)$ denotes the Gamma function \cite{mason1995optical, mason2000estimating}. The real and imaginary parts of $|G^{*}_{\omega_j}|$ correspond to the 
storage modulus $G'_{\omega_j}$ and the loss modulus $G''_{\omega_j}$, given by
\begin{align}
G'_{\omega_j} = |G^{*}_{\omega_j}| \cos\,\left(\tfrac{\pi {\alpha}_{\Delta t_j}}{2}\right) \label{equ:Gp}\\ 
G''_{\omega_j} = |G^{*}_{\omega_j}| \sin\,\left(\tfrac{\pi {\alpha}_{\Delta t_j}}{2}\right).
\label{equ:Gpp}
\end{align}

Both quantities depend on the values of MSD and the corresponding log slope, and therefore rely largely on accurate estimation of the slope vector $\bm \alpha = [\alpha_{\Delta t_1}, \dots, \alpha_{\Delta t_{n-1}}]^T$. In MF-AIUQ, we approximate this value through finite differences of the estimated MSD curve. For interior points $j = 2,\, \dots, \,n-2$, 
\begin{equation}
    \alpha_{est,\Delta t_j} \approx \frac{\tilde \theta_{est,\Delta t_{j+1}} - \tilde \theta_{est,\Delta t_{j-1}}}{\log \Delta t_{j+1} - \log \Delta t_{j-1}},
    \label{equ:alpha_MF_AIUQ_1}
\end{equation}
At the endpoints, use one-sided differences, with 
\begin{align}
    & \alpha_{est,\Delta t_1} \approx \frac{\tilde \theta_{est,\Delta t_2} - \tilde \theta_{est,\Delta t_1}}{\log \Delta t_{2} - \log \Delta t_{1}},\label{equ:alpha_MF_AIUQ_2}\\
    & \alpha_{est,\Delta t_{n-1}} \approx \frac{\tilde \theta_{est,\Delta t_{n-1}} - \tilde \theta_{est,\Delta t_{n-2}}}{\log \Delta t_{n-1} - \log \Delta t_{n-2}}.
    \label{equ:alpha_MF_AIUQ_3}
\end{align}
The storage and loss moduli are then derived along with their corresponding frequencies. 
{Because the transformation from MSD to modulus through the GSER is nonlinear, applying the predictive mean of GSER directly to point estimates of the MSD may not lead to the predictive mean of the distributions of the storage and loss moduli. Therefore, instead of using only the point estimates of the MSD, we propagate MSD uncertainty through Monte Carlo sampling. The sampled log MSD curves are transformed through the GSER to obtain samples of the storage and loss moduli, and the pointwise medians of these samples are used as the final estimates. Details are provided in} Appendix \ref{app:MSD_Gp_Gpp} with the procedure summarized in Algorithm \ref{alg:gser_moduli}.

One challenge in analyzing the data obtained using MPT is that the log MSD curve is often not smooth and exhibits substantial local fluctuations, prohibiting directly estimating the derivative of MSD by finite difference. To address this challenge, we consider two smoothing strategies. In the first approach, we apply cubic smoothing spline regression \cite{reinsch1967smoothing} to log MSD obtained from MPT as a function of $\log \Delta t$, implemented in MATLAB using the \texttt{csaps} function. The resulting smooth and differentiable curve is then used to estimate the logarithmic slope $\bm \alpha$ using Eqs. (\ref{equ:alpha_MF_AIUQ_1}), (\ref{equ:alpha_MF_AIUQ_2}), and (\ref{equ:alpha_MF_AIUQ_3}). In the second approach, log MSD is approximated by a fourth-order polynomial in $\log \Delta t$, and $\bm\alpha$ is obtained by differentiation \cite{gu2021uncertainty}. The pointwise standard deviations of the MSD can be obtained from MPT, yet the error for smoothing MSD curves is hard to propagate. Therefore, we directly apply the GSER to the smoothed MSD curve to estimate the storage and loss moduli, rather than using sample medians as in MF-AIUQ in the numerical study presented in Section \ref{sec:experiment} D.

\section{\label{sec:simulation}SIMULATION STUDIES}
\begin{figure*}[t]
    \centering
    \begin{tabular}{ccc}
    \multicolumn{3}{c}{
        \begin{overpic}[scale=0.55]{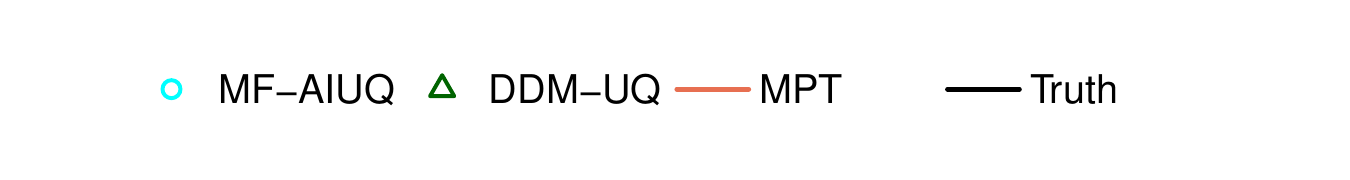}
        \end{overpic}
    } \\[-2em]
    \begin{overpic}[scale=0.4]{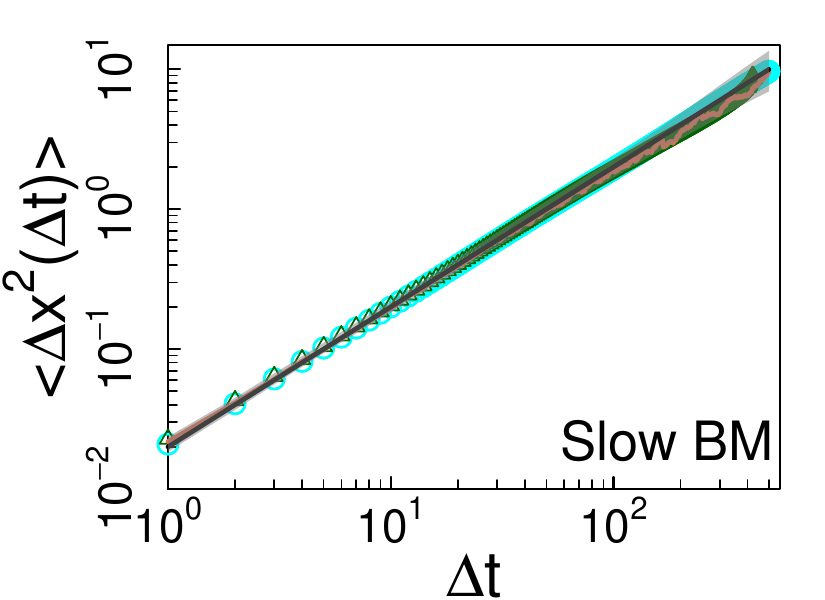}
            \put(0.5,64){\textbf{(a)}}
        \end{overpic} &
        \begin{overpic}[scale=0.4]{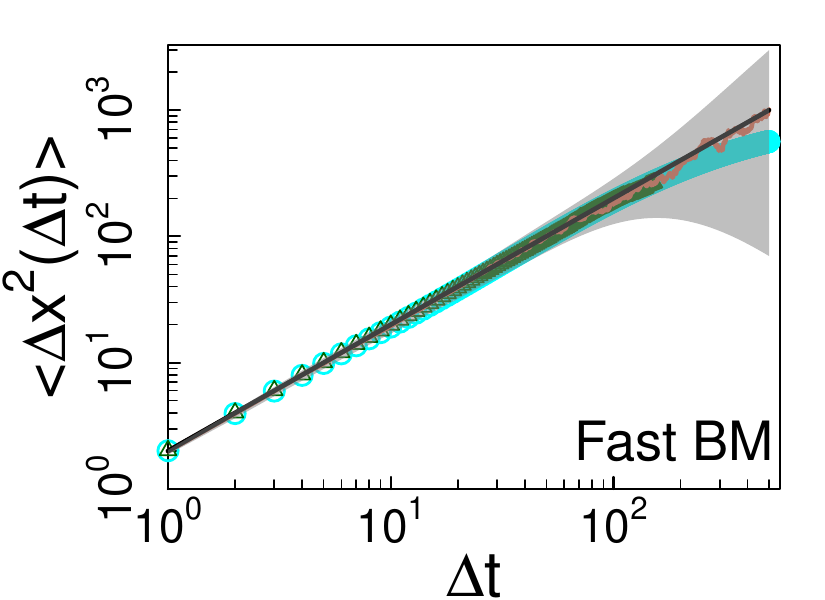}
            \put(0.5,64){\textbf{(b)}}
        \end{overpic} &
         \begin{overpic}[scale=0.4]{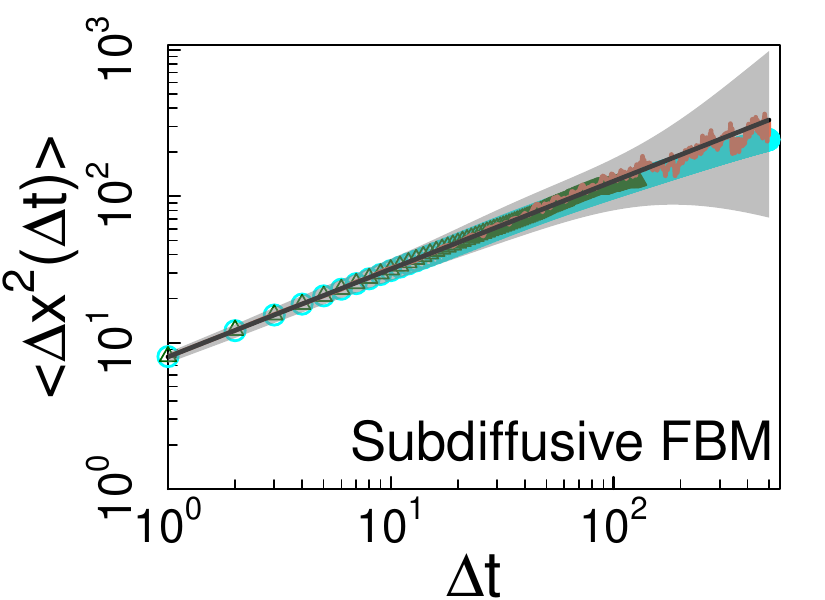}
            \put(0.5,64){\textbf{(c)}}
        \end{overpic} \\
        \begin{overpic}[scale=0.4]{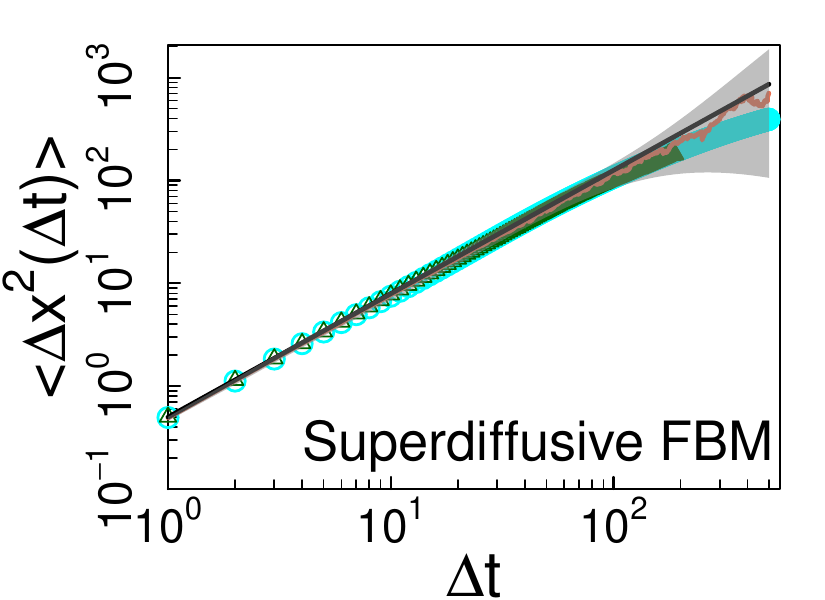}
            \put(0.5,64){\textbf{(d)}}
        \end{overpic} &
        \begin{overpic}[scale=0.4]{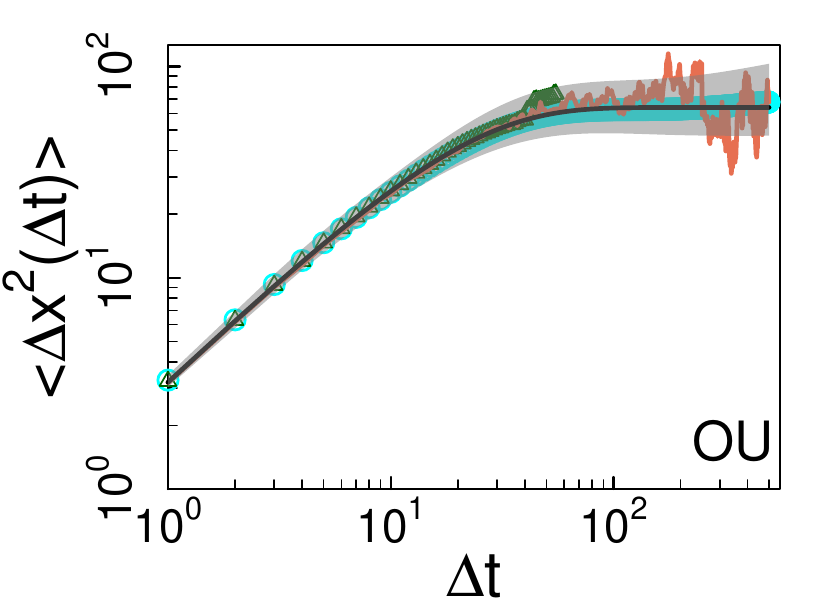}
            \put(0.5,64){\textbf{(e)}}
        \end{overpic} &
         \begin{overpic}[scale=0.4]{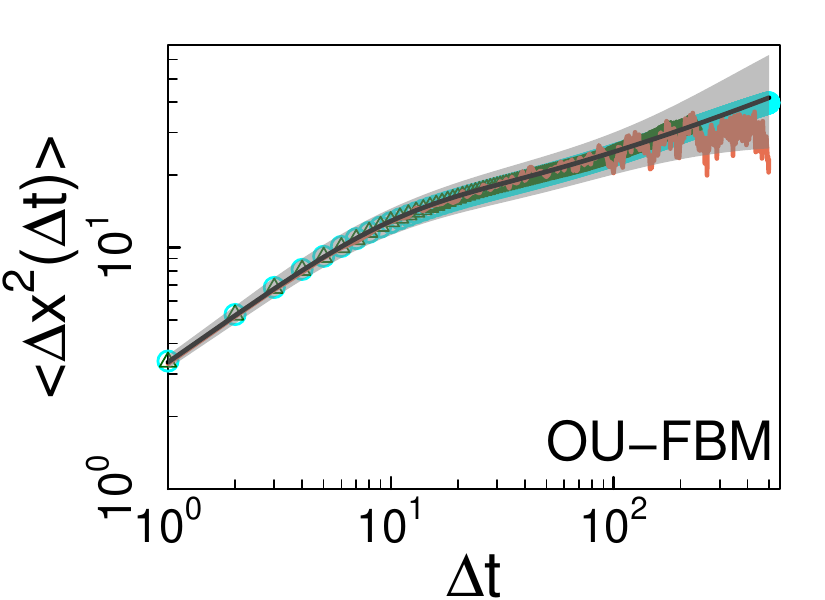}
            \put(0.5,64){\textbf{(f)}}
        \end{overpic}
    \end{tabular}
    \caption{MSD versus lag time for $M=100$ simulated particles. MSD values are estimated using MF-AIUQ (cyan circles), DDM-UQ (dark green triangles), and MPT (orange solid lines). The theoretical MSD curves are denoted by black solid lines. The six panels correspond to (a) slow Brownian motion with $\sigma^2_{BM} = 0.02$, (b) fast Brownian motion with $\sigma^2_{BM} = 2$, (c) subdiffusive fractional Brownian motion with $\sigma^2_{FBM} = 8$, $\alpha = 0.6$, (d) superdiffusive fractional Brownian motion with $\sigma^2_{FBM} = 0.5$, $\alpha = 1.2$, (e) Ornstein–Uhlenbeck process with $\sigma^2_{OU} = 64$, $\rho = 0.95$, and (f) mixture of OU and FBM with $\sigma^2_{OU} = 9$, $\rho = 0.85$, $\sigma^2_{FBM} = 2$, and $\alpha = 0.45$. The shaded regions denote the $95\%$ confidence intervals of MSD estimates from MF-AIUQ.}
    \label{fig:simulation_MSD}
\end{figure*}

We first conduct simulation studies in which particle trajectories are generated from stochastic processes with known theoretical MSD values. All processes are assumed to be isotropic, so that the dynamics are identical in all spatial directions. We consider simulation settings designed to cover several representative dynamical behaviors. Specifically, six videos are simulated, each containing 100 particle trajectories on a $500\times 500$ pixel grid over 500 time frames, with minimum lag time $\Delta t_{min} = 1$. The simulations include two BM processes, with $\sigma^2_{BM} = 0.02$ for slow dynamics and $\sigma^2_{BM} = 2$ for fast dynamics, two fractional Brownian motion (FBM), with $\sigma^2_{FBM} = 8$ and $\alpha = 0.6$ for subdiffusive behavior, and $\sigma^2_{FBM} = 0.5$, $\alpha = 1.2$ for superdiffusive behavior, one OU process with $\sigma^2_{OU} = 64$ and $\rho = 0.95$, and a mixture of OU and FBM (OU-FBM) processes with $\sigma^2_{OU} = 9$, $\rho = 0.85$, $\sigma^2_{FBM} = 2$, and $\alpha = 0.45$. Further details of the simulation are provided in Appendix \ref{app:simulation}. The computational cost of MF-AIUQ is provided in Fig. \ref{fig:computation_time} in Appendix \ref{app:computational_cost}, which costs less than 1 minute for processing one simulation video of $500\times 500 \times 500$ on a desktop computer.

We compare MF-AIUQ with two existing model-free approaches, DDM-UQ and MPT. All three methods estimate the MSD directly without assuming a known parametric model for the underlying dynamics. Compared with the MD-AIUQ approach of \cite{gu2024ab}, this setting is more flexible but also more challenging. The MF-AIUQ estimates are obtained using Algorithm \ref{alg:MF_AIUQ} with the default settings for all scenarios. For MPT, we use a common parameter setting across five of the six simulations to keep the comparison consistent. In the subdiffusive FBM case, however, this common setting provides nonzero MSD estimates over less than half of the full lag time range, with substantial variability at large lag times. We therefore use a case-specific parameter setting to improve the estimation of MPT in this simulation. This illustrates the sensitivity of MPT to user-specified tracking parameters, whereas MF-AIUQ does not require such tuning.

The resulting MSD estimates are compared with the true MSD values to assess estimation accuracy, and are shown in Fig. \ref{fig:simulation_MSD}. MF-AIUQ estimates are denoted by cyan circles, DDM-UQ by dark green triangles, MPT by orange solid lines, and the true MSD  used for simulation by the black solid lines.

Across the six dynamical processes, MF-AIUQ accurately estimates the overall shape and magnitude of the true MSD curves over most of the lag time range. Compared with MPT, MF-AIUQ produces smoother MSD estimates, while the MPT estimates exhibit increasingly unstable estimation at large lag times, particularly in the OU and mixed OU-FBM cases. For all methods, MSD estimation is generally more accurate at small lag times than at large lag times, as more information is available in the small lag region. However, DDM-UQ only provides estimation over a limited range of lag times, whereas MF-AIUQ provides estimates over the full range. Furthermore, the $95\%$ confidence intervals of the MSD  from MF-AIUQ cover most of the true MSD curves, indicating appropriate uncertainty quantification, and the wider confidence intervals at larger lag times reflect the increased estimation uncertainty due to limited information in the tail region of the MSD curve.

To quantify estimation accuracy, we compute the root mean squared error (RMSE) of the MSD estimates from MF-AIUQ and MPT. DDM-UQ is excluded from this comparison as it provides estimates over only a limited range of lag times. We also provide the standard deviation (SD) of the theoretical MSD values, which serves as a reference for the overall variation of the true MSD values. Both RMSE and SD are computed on the logarithmic scale with base 10 for consistency with the figures:
\begin{align*}
  & \mathrm{RMSE} = \sqrt{\frac{1}{n-1} \sum_{j=1}^{n-1}(\log_{10}\theta_{est,\Delta t_j} - \log_{10}\theta_{true, \Delta t_j})^2}, \\
  & SD = \sqrt{\frac{1}{n-2}\sum_{j=1}^{n-1}(\log_{10}\theta_{true,\Delta t_j} - m_{\log_{10}\theta})^2}.
\end{align*}
Here, $\theta_{est, \Delta t_j}$ and $\theta_{true, \Delta t_j}$ represent the estimated and true MSD at lag time $\Delta t_j$, for $j = 1, \dots, n-1$, and $m_{\log_{10}\theta}$ represents the mean of the true MSD on the $\log_{10}$ scale across all lag times. Details on the calculation of the true MSD are provided in Appendix \ref{app:simulation}. 

\begin{table}[htb]
\centering
\begin{tabular}{lcccc}
\hline
Simulation & MF-AIUQ & MPT & SD \\
\hline
Slow BM             & \textbf{0.0109} & 0.0532 & 0.4238 \\
Fast BM             & 0.1369 & \textbf{0.0403} & 0.4238 \\
Subdiffusive FBM   & 0.0887 & \textbf{0.0452} & 0.2543 \\
Superdiffusive FBM & 0.2074 & \textbf{0.0784} & 0.5086 \\
OU                  & \textbf{0.0116} & 0.1050 & 0.1123 \\
OU-FBM            & \textbf{0.0137} & 0.0929 & 0.1347 \\
\hline
\end{tabular}
\caption{RMSE of MSD estimates for simulated data, computed using the logarithm with base 10, together with the SD of the true MSD. The lowest RMSE in each case is shown in bold.}
\label{tab:simulation_rmse}
\end{table}

The RMSE results are summarized in Table \ref{tab:simulation_rmse}. Both MF-AIUQ and MPT have RMSE values smaller than the SD of the true MSD, indicating that both methods provide sensible estimation. In the OU and OU-FBM settings, where the MSD curves flatten at long lag times, the RMSE values of MPT, however, are close to the SD of the true MSD curves. Among the six scenarios, MF-AIUQ has a smaller RMSE than MPT for slow BM, OU, and OU-FBM, while MPT provides better estimates for the other three simulations.

Overall, the simulation studies show that MF-AIUQ provides an accurate model-free estimation of MSD at the entire time lag range from microscopy videos, which does not require parameter tuning and the estimated MSD curves are typically smoother than those from MPT, particularly at larger time points. In the following section, we consider three experimental settings to further demonstrate that MF-AIUQ serves as a useful complementary approach to MPT for MSD estimation when trajectory reconstruction is difficult, and to MD-AIUQ when the particle dynamical model is inaccessible.

\section{\label{sec:experiment}REAL EXPERIMENTAL ANALYSIS}
\subsection{Materials and Methods}

Snail mucin was purchased from COSRX Inc. (Gangnam-gu, Seoul, Republic of Korea). The raw material contains 96\% snail mucin secretion and a small amount of additives. It is diluted with deionized water in a 1:1 ratio and otherwise used without modification. Diluted mucin is mixed with tracer particles ($r$ = 250 nm), and a thin layer ($\sim$ 100 $\mu$L) was pipetted into a 35 mm glass-bottom Petri dish (MatTek Corporation, Ashland, MA) and sealed with a layer of oil to minimize evaporation. Images were acquired by an inverted fluorescence microscope (Zeiss Axio Observer 7, White Plains, NY) with a 40$\times$ objective in the GFP channel (peak wavelength = 517 nm) at a frame rate of 3.33 per second with a size of 512 $\times$ 512 pixels and a spatial resolution of $0.1465\, \mu\text{m/pixel}$, for 2000 frames. The experiment is conducted at a temperature $T_a = 293\, K$.

To obtain reference storage and loss moduli, bulk rheometry of snail mucin was performed using a TA Instruments Discovery HR-30 rheometer (TA Instruments, New Castle, DE) equipped with a 40 mm parallel-plate geometry and a sand-blasted Peltier lower plate. $\sim$900 $\mu$L of the mucin solution was loaded onto the lower plate using a pipette, and the gap thickness was fixed at 700 $\mu$m. Oscillatory frequency sweeps were conducted over angular frequencies from 100 to $10^{-3}$ $\mathrm{rad/s}$, with five points per decade, at a fixed strain amplitude of $0.5\%$.

The materials of PVA solution and tertraPEG hydrogels experiments studied in Section \ref{subsec:pva} and Section \ref{subsec:hydrogel} were introduced in Ref. \cite{gu2024ab} and the experiments are analyzed by the MD-AIUQ approach based on pre-specified parametric forms of MSD curve. We re-analyze these experiments by the MF-AIUQ developed in this article.

\subsection{MSD estimation for tracer probes of different sizes in a polymer solution}
\label{subsec:pva}


\begin{figure*}[t]
    \centering
    \begin{tabular}{ccc}
    \multicolumn{3}{c}{
        \begin{overpic}[scale=0.5]{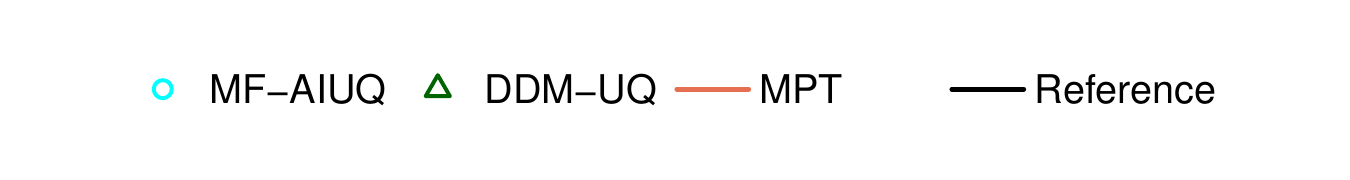}
        \end{overpic}
    } \\[-2em]
    \begin{overpic}[scale=0.35]{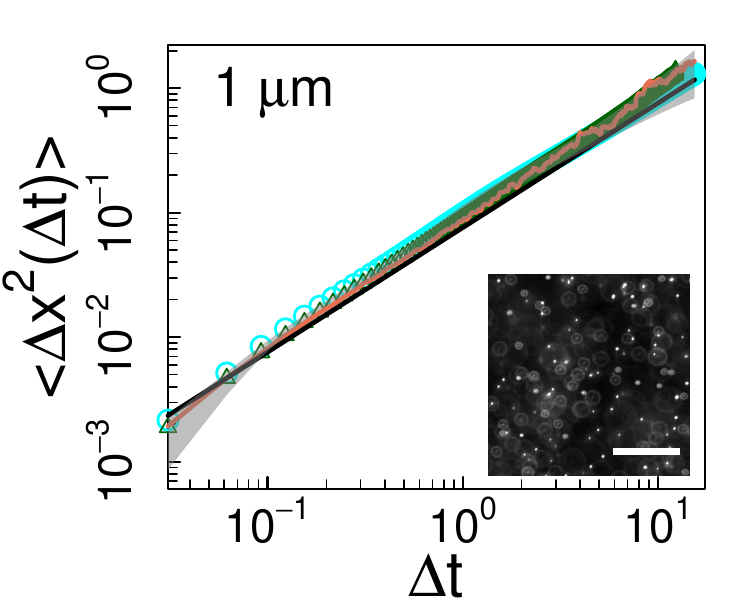}
            \put(0.5,72){\textbf{(a)}}
        \end{overpic} &
        \begin{overpic}[scale=0.35]{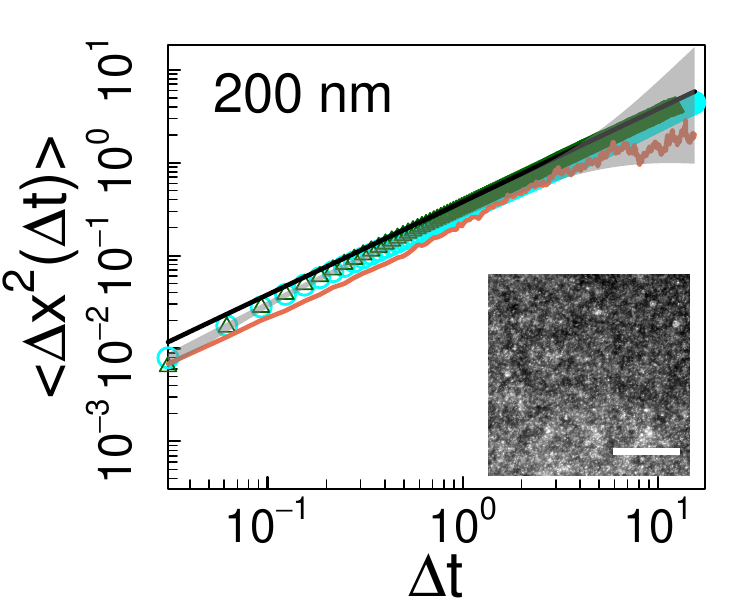}
            \put(0.5,72){\textbf{(b)}}
        \end{overpic} &
         \begin{overpic}[scale=0.35]{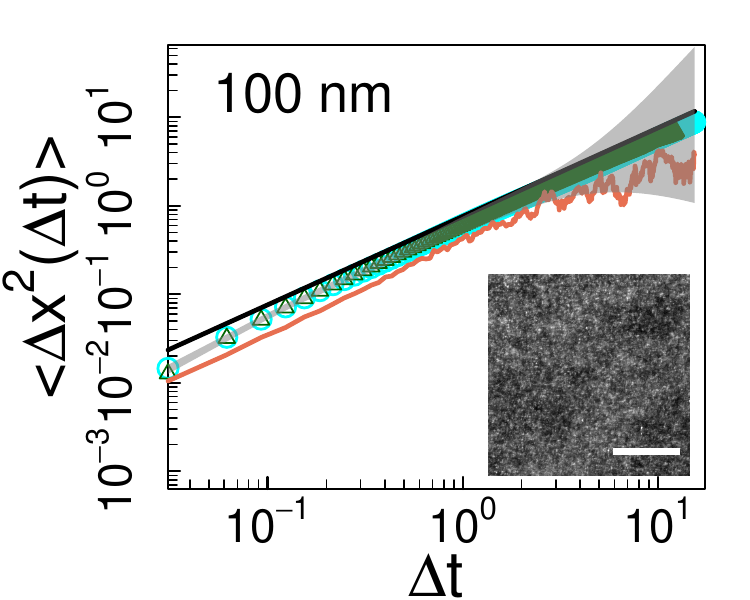}
            \put(0.5,72){\textbf{(c)}}
        \end{overpic} 
        \begin{overpic}[scale=0.35]{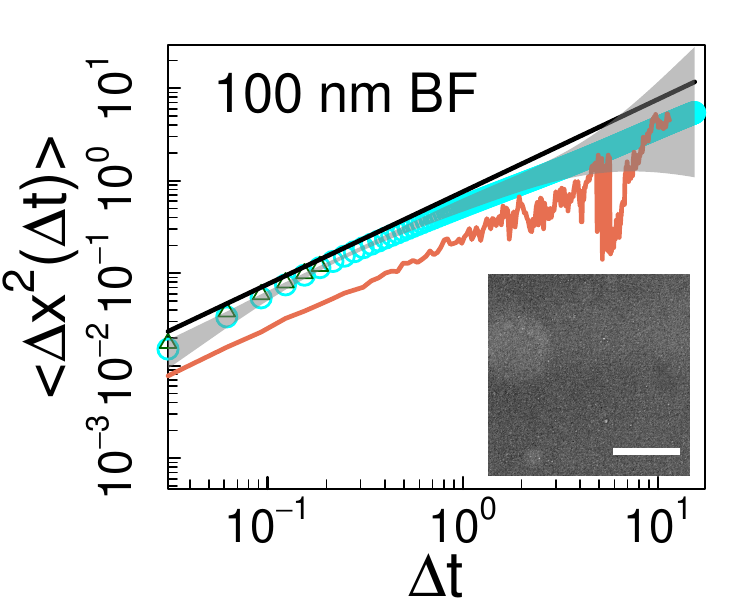}
            \put(0.5,72){\textbf{(d)}}
        \end{overpic}
    \end{tabular}
    \caption{MSD versus lag time of probes in a 4 (w/v)\% PVA solution. The embedded particle sizes are: (a) $2r$ = $1\, \mu m$, (b) $2r$ = $200\ nm$, (c) $2r$ = $100\ nm$, and (d) $2r$ = $100\ nm$ under bright-field imaging. The estimation uses MF-AIUQ (cyan circles), DDM-UQ (dark green triangles), and MPT (orange solid lines). The reference MSD curves are denoted by the solid black lines. The shaded regions denote the $95\%$ confidence intervals from MF-AIUQ. Representative images from the corresponding videos are shown in the bottom right insets. Scale bars represent $50\,\mu m$.}
    \label{fig:real_eg_1_4pct_PVA_MSD}
\end{figure*}

We first consider experimental microscopy data of tracer particles embedded in a 4~(w/v)\% polyvinyl alcohol (PVA) solution. Image sequences of size $512 \times 512$ pixels, with a resolution of 0.29 $\mu$m/pixel, were recorded over $n = 500$ time frames with time step size of $\Delta t_{min} = 0.0309$ seconds. The tracer probes have diameters of $2r$ = 1 $\mu$m, 200 nm, and 100 nm. Four experiments were conducted, including three fluorescence imaging experiments using tracer particles of these sizes, and an additional experiment using 100 nm particles under bright-field imaging. All experimental details regarding sample preparation and imaging are provided in Ref. \cite{gu2024ab}.

The PVA solution behaves approximately as a Newtonian fluid with viscosity $\eta \approx 25$ mPa$\cdot$s, and thus a reference MSD can be computed from the Stokes-Einstein relation:
\begin{equation}
\theta_{\Delta t}
= \frac{2 k_B T_a}{3 \pi \eta r}\,\Delta t,
\label{equ:Stokes-Einstein}
\end{equation}
where $k_B = 1.38\times 10^{-23}\,J/K$ denotes the Boltzmann constant, $T_a$ is the absolute temperature, and $r$ represents the particle radius \cite{gu2024ab}. 

According to Eq. (\ref{equ:Stokes-Einstein}), the MSD increases as the particle radius decreases when the temperature and viscosity are held constant. Particle size also affects the difficulty of particle tracking. Although all samples have the same volume fraction, smaller particles correspond to a larger number of probes and higher optical density, making particle identification and trajectory linking more challenging for MPT \cite{bayles2017probe}. The 100 nm bright-field video introduces an additional challenge because the reduced contrast relative to the background further hinders reliable particle identification \cite{edera2017differential}. Thus, this experimental setting allows us to assess the robustness of MF-AIUQ under challenging imaging conditions. We highlight that model-free estimation by MF-AIUQ tackles a much harder problem than the one in Ref. \cite{gu2024ab}, where MD-AIUQ assumes diffusive dynamics and estimates only the diffusion coefficient, while the process is not assumed to be diffusive in MF-AIUQ \textit{a priori}.

MSD estimates from MF-AIUQ, DDM-UQ, and MPT, together with the theoretical reference from the Stokes-Einstein relation, are shown in Fig.~\ref{fig:real_eg_1_4pct_PVA_MSD}. Under fluorescence imaging, estimates from all three methods align well with the reference MSD for the 1 $\mu$m particles. 
For the 200 nm and 100 nm particles, MPT underestimates the MSD over all lag times, as the higher optical density makes particle identification and trajectory linking less reliable. In the 100 nm bright-field case, the MPT curve lies below the reference over the entire lag time range, which we attribute to the low signal-to-noise ratio of the images, and DDM-UQ provides estimates consistent with the reference only over a limited range. 
In contrast, the estimates from MF-AIUQ remain close to the reference over the full lag time range in all four cases, with a default, tune-free setting, same as the one from the simulation.

\begin{table}[!htbp]
\centering
\begin{tabular}{lcccc}
\hline
Particle size (2$r$) & MF-AIUQ & MPT & SD \\
\hline
$1\ \mu m$    & \textbf{0.0711} & 0.1314 & 0.4238 \\
$200\ nm$     & \textbf{0.1222} & 0.3426 & 0.4238 \\
$100\ nm$     & \textbf{0.1194} & 0.4490 & 0.4238 \\
$100\ nm$ BF  & \textbf{0.2675} & 0.6003 & 0.4211 \\
\hline
\end{tabular}
\caption{RMSE of MSD estimates for probes in a 4 (w/v)\% PVA solution, computed using the logarithm with base 10, together with the SD of the reference MSD. The lowest RMSE in each case is shown in bold.
}
\label{tab:real_eg_1_4pct_PVA_rmse}
\end{table}

The RMSE values for MF-AIUQ and MPT are shown in Table \ref{tab:real_eg_1_4pct_PVA_rmse}. Across all four experimental cases, MF-AIUQ has smaller RMSE values than MPT. For the high density systems, the RMSE value of MPT is close to the SD of the reference MSD for the $200$ nm probes and exceeds the SD for the $100$ nm probes. In the bright-field experiment, which also uses $100$ nm probes, the RMSE of MPT is substantially larger than the SD. These results indicate that MF-AIUQ remains reliable in extracting dynamical information even under challenging imaging conditions such as high particle density or low image contrast.

\subsection{Particle dynamics during gelation of tetraPEG hydrogel}
\label{subsec:hydrogel}

\begin{figure*}[!t]
    \centering
    \begin{tabular}{ccc}
    \multicolumn{3}{c}{
        \begin{overpic}[scale=0.55]{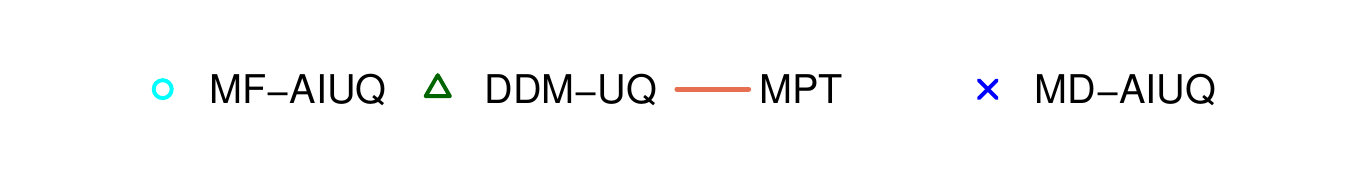}
        \end{overpic}
    } \\[-2em]
    \begin{overpic}[scale=0.4]{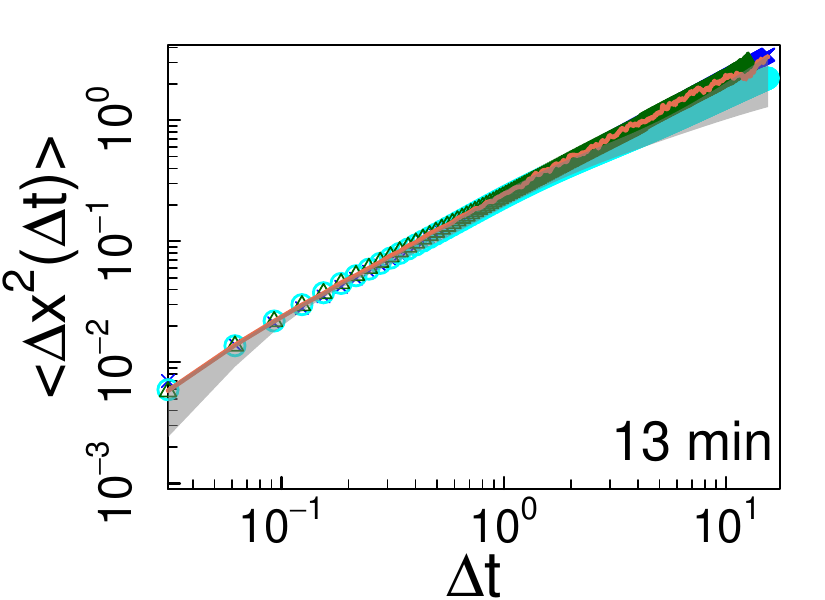}
            \put(0.5,64){\textbf{(a)}}
        \end{overpic} &
        \begin{overpic}[scale=0.4]{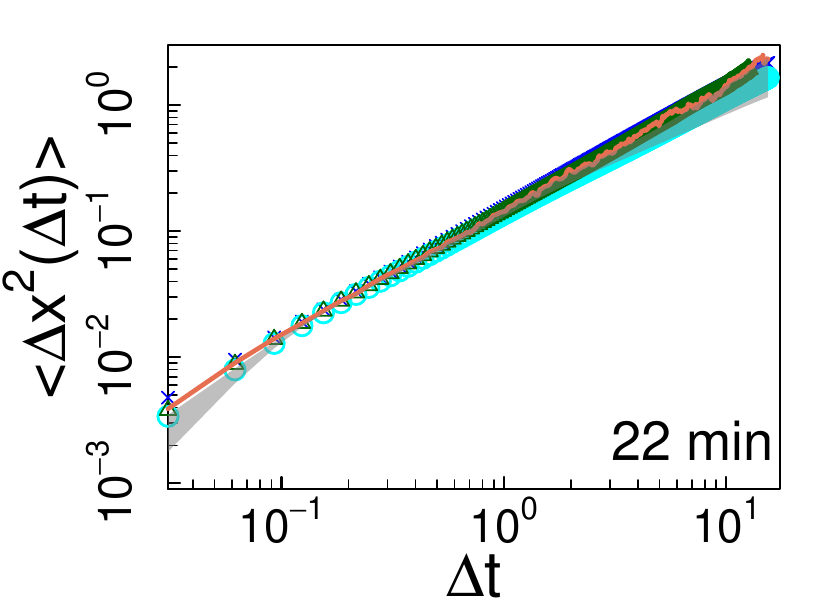}
            \put(0.5,64){\textbf{(b)}}
        \end{overpic} &
         \begin{overpic}[scale=0.4]{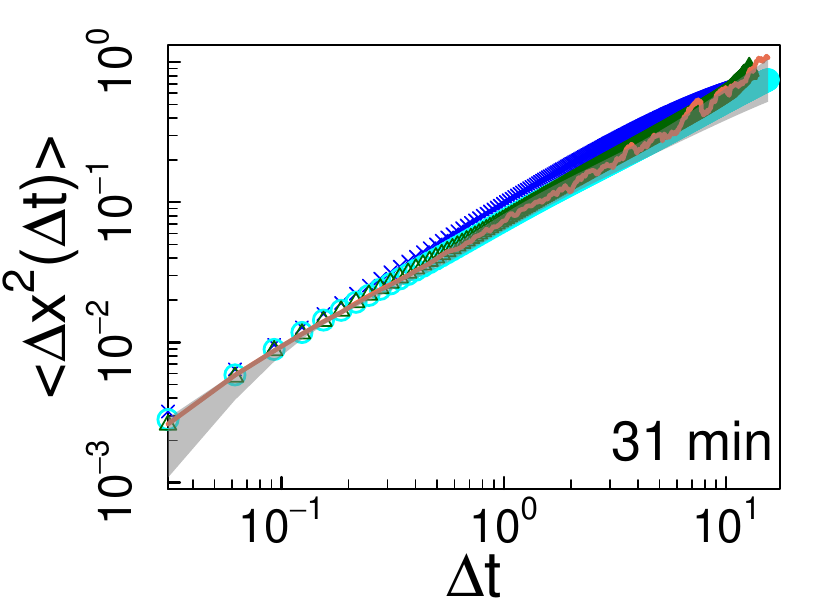}
            \put(0.5,64){\textbf{(c)}}
        \end{overpic} \\
        \begin{overpic}[scale=0.4]{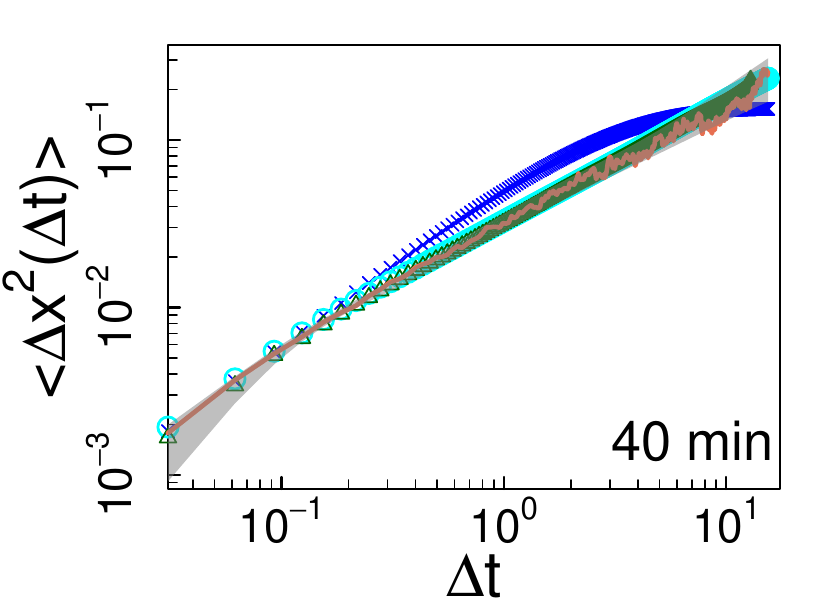}
            \put(0.5,64){\textbf{(d)}}
        \end{overpic} &
        \begin{overpic}[scale=0.4]{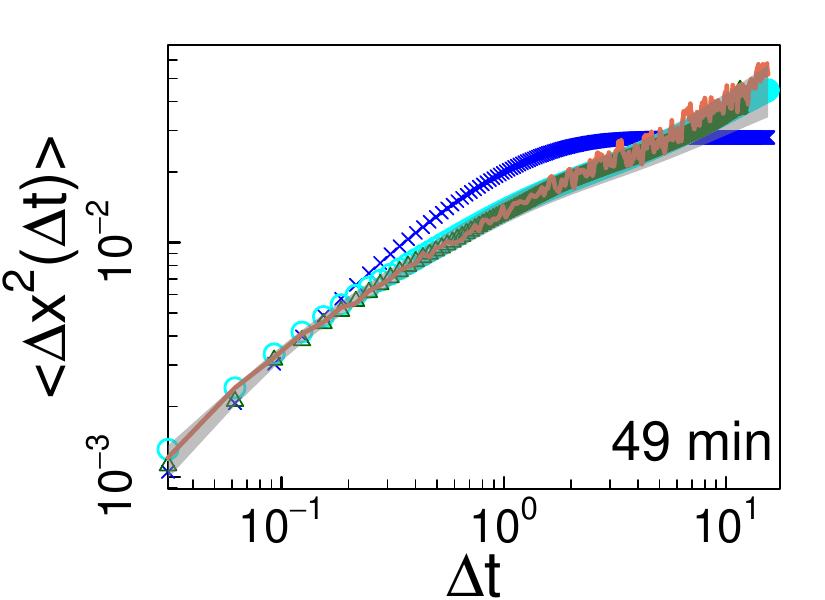}
            \put(0.5,64){\textbf{(e)}}
        \end{overpic} &
         \begin{overpic}[scale=0.4]{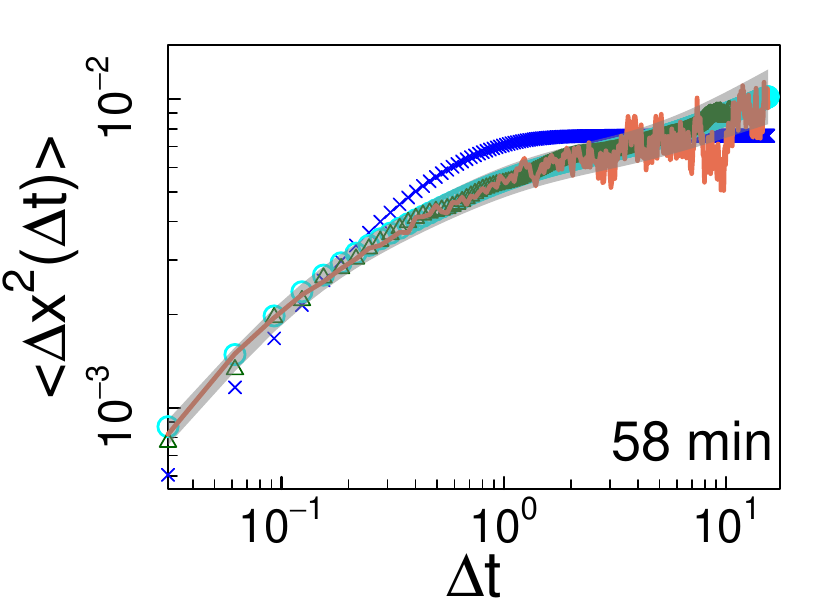}
            \put(0.5,64){\textbf{(f)}}
        \end{overpic}
    \end{tabular}
    \caption{MSD versus lag time of PEG-hydrogel data. The estimation uses MF-AIUQ (cyan circles), MD-AIUQ (blue crosses), MPT (orange solid lines), and DDM-UQ (dark green triangles). The shaded regions denote the $95\%$ confidence intervals of the MSD estimates from MF-AIUQ.}
    \label{fig:real_eg_2_PEG_Hydrogel_MSD}
\end{figure*}

\begin{figure*}[!t]
    \centering
    \begin{tabular}{cc}
    \begin{overpic}[scale=0.52]{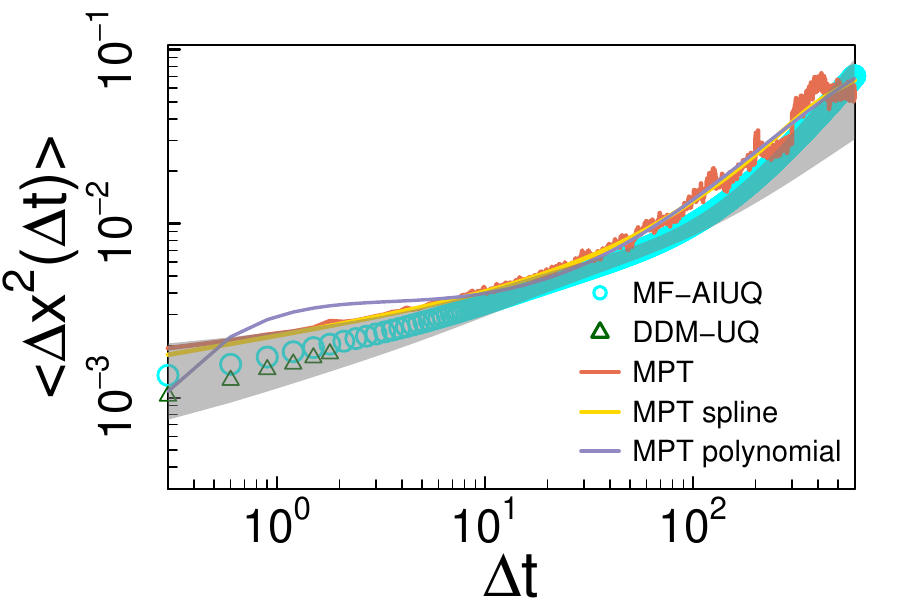}
            \put(0.5,64){\textbf{(a)}}
        \end{overpic} &
        \begin{overpic}[scale=0.52]{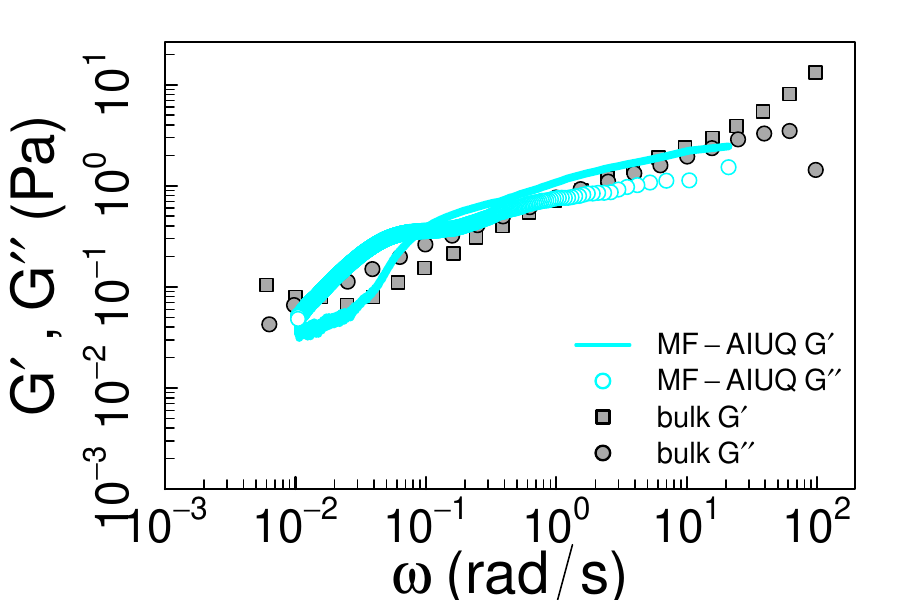}
            \put(0.5,64){\textbf{(b)}}
        \end{overpic} \\
         \begin{overpic}[scale=0.52]{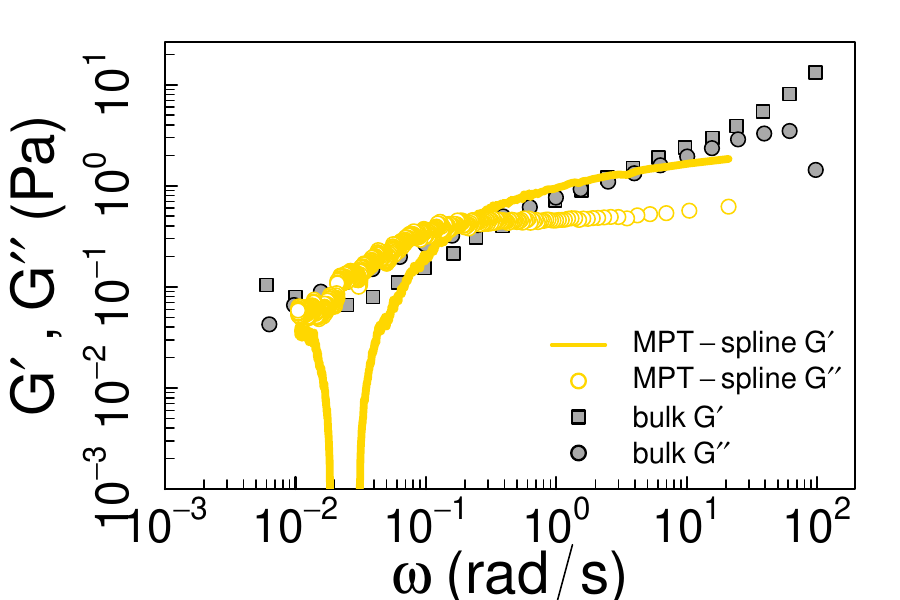}
            \put(0.5,64){\textbf{(c)}} 
        \end{overpic} &
          \begin{overpic}[scale=0.52]{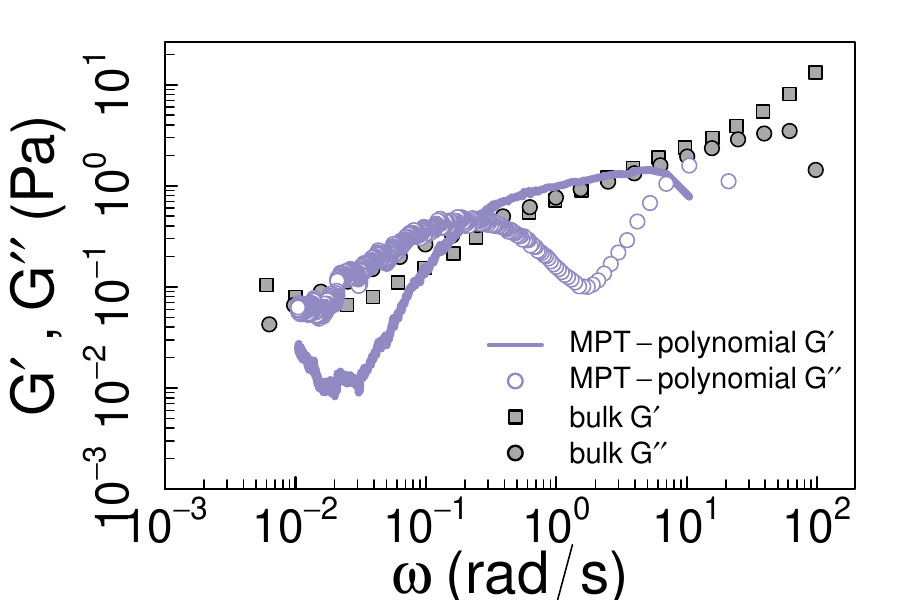}
            \put(0.5,64){\textbf{(d)}}
             \end{overpic} 
    \end{tabular}
    \caption{Results of microrheology and bulk rheology measurements of snail mucin. (a) MSD obtained using MF-AIUQ (cyan circles), MPT (orange solid lines), and DDM-UQ (dark green triangles). The gold solid line represents the spline fitted MSD from MPT. The purple solid line represents the polynomial fitted MSD from MPT. The shaded regions denote the $95\%$ confidence intervals of MSD estimates from MF-AIUQ. (b)-(d) Comparison of the frequency dependent linear viscoelastic moduli from bulk rheology experiments with those computed from the estimated MSDs. Bulk measurements are shown as gray squares for $G'_{\omega}$ and gray circles for $G''_{\omega}$. Panel (b) shows the MF-AIUQ estimates in cyan, with $G'_{\omega}$ plotted as a solid line and $G''_{\omega}$ as hollow circles. Panel (c) shows the MPT estimates based on the spline smoothed MSD in orange, with $G'_{\omega}$ plotted as a solid line and $G''_{\omega}$ as hollow circles. Panel (d) shows the MPT estimates based on the polynomial approximated MSD in orange, with $G'_{\omega}$ plotted as a solid line and $G''_{\omega}$ as hollow circles.}
    \label{fig:real_eg_3_MSD_Gp_Gpp}
\end{figure*}

Gelation is a process in which a material undergoes a gradual transition from fluid-like to solid-like behavior \cite{larsen2008microrheology}. During this transition, tracer particle dynamics become increasingly constrained, and the MSD curves change from diffusive behavior towards a plateau at longer lag times. Here, we use the process of gelation to assess whether MF-AIUQ can estimate time-dependent particle dynamics without imposing a parametric model as in MD-AIUQ.

We analyze the gelation of tetraPEG hydrogel capped with functional groups succinimidyl glutarate (-SG) and amine (-NH$_2$) with a stoichiometric ratio of SG:NH$_2$ = 1:1, following the experimental setting studied in Refs. \cite{parrish2020nanoparticle, gu2024ab,luo2025optimizing}. Particle dynamics are examined at six selected times, which span the transition from early fluid-like stages to later more constrained dynamics, at 13, 22, 31, 40, 49, and 58 minutes after mixing the two tetraPEGs.

The estimated curves show a systematic decrease in logarithmic slope over experimental time, with more significant flattening at long lag times in the later stages (Fig. \ref{fig:real_eg_2_PEG_Hydrogel_MSD}). This pattern is consistent with the previous studies on gelation \cite{parrish2020nanoparticle, gu2024ab}. Although no ground truth MSD is available in this experiment, the MSD estimates from MF-AIUQ and MPT are generally consistent across the six experiments, with MF-AIUQ showing smoother curves throughout the entire lag time range. The smooth MSD curves by MF-AIUQ can facilitate further analyses such as automated data superposition for gelation point identification, where non-smooth MPT estimates may require additional preprocessing \cite{lennon2023data}. The DDM-UQ estimates are close to those from MF-AIUQ and MPT at small lag times, but are not available at larger lag times because the estimation becomes unstable.

In the previous study \cite{gu2024ab}, MD-AIUQ assumes an OU process and has been shown to accurately identify the gelation point in an automated manner. The MD-AIUQ estimates are closer to the model-free estimates at the initial experimental times, and previous results suggest similar agreement at the final experimental times \cite{gu2024ab}, but larger discrepancies appear in the intermediate stages, where the underlying dynamic model is more difficult to specify. The difference between MD-AIUQ and the model-free estimates suggests that the OU process assumed in MD-AIUQ may not be flexible enough to capture the MSD shape across all stages of gelation. This pattern reflects a trade-off between model interpretability and flexibility. Although MD-AIUQ can detect dynamic transitions to gelation, MF-AIUQ is better at capturing the evolving MSD curves when the underlying parametric model is unknown.

\subsection{Microrheological estimation of storage and loss moduli in snail mucin}

Snail mucin is a viscoelastic biological material with applications in biomaterials, biomedical engineering, and skincare \cite{deng2023natural, mcdermott2021advancing}. Measuring its properties is important for quality control and for performance optimization in practical applications. 
In our experiment, snail mucin was diluted 1:1 with DI water to provide sufficient tracer mobility for microrheological measurements while preserving a measurable response in bulk rheometry. After dilution, the sample appeared macroscopically homogeneous, with no visible phase separation or large aggregates. Microrheology measurements from different fields of view showed consistent tracer dynamics.
This experiment evaluates the performance of MF-AIUQ in estimating storage and loss moduli, using bulk rheology measurements as a reference.

MSD estimates from MF-AIUQ, DDM-UQ, and MPT are shown in Fig. \ref{fig:real_eg_3_MSD_Gp_Gpp}(a). MF-AIUQ and MPT show similar trends over most lag times, with the main difference appearing at short lag times, where the MF-AIUQ curve lies below the MPT curve. DDM-UQ provides estimates only at the first six lag times, and these estimates are close to those from MF-AIUQ.

We then use the MSD values to find the storage and loss moduli from MF-AIUQ and MPT. For MF-AIUQ, the moduli are obtained using Algorithm \ref{alg:gser_moduli}, where the pointwise medians of the $G'_{\omega}$ and $G''_{\omega}$ samples are used as the estimates. For MPT, we use the two smoothing strategies described in Section \ref{sec:storage_loss_modulus} before moduli estimation. The resulting moduli with the bulk rheology measurements are shown in Fig. \ref{fig:real_eg_3_MSD_Gp_Gpp}(b)-(d), where panel (b) shows the moduli from MF-AIUQ and panels (c) and (d) show the moduli from MPT using spline smoothing and polynomial approximation, respectively. Overall, the viscoelastic moduli derived from MF-AIUQ show closer agreement with the bulk measurements compared with those derived from MPT, as shown in Fig. \ref{fig:real_eg_3_MSD_Gp_Gpp}(b).

The $G'_{\omega}$ estimate first decreases and then rises sharply at low frequencies (Fig. \ref{fig:real_eg_3_MSD_Gp_Gpp}(c)). This behavior is associated with the tail of the smoothed MSD curve of MPT, shown as the gold line in Fig. \ref{fig:real_eg_3_MSD_Gp_Gpp}(a), where the slope changes slightly at large lag times. As a result, the estimated $\alpha_{\Delta t}$ increases above 1 before decreasing below 1 again. According to Eq. (\ref{equ:Gp}), values of $\alpha_{\Delta t} > 1$ lead to negative estimates of $G'_{\omega}$. The polynomial approximation of MPT in Fig. \ref{fig:real_eg_3_MSD_Gp_Gpp}(d) does not exhibit the same issue at low frequencies, yet the resulting moduli deviate significantly from the bulk rheology measurements. This discrepancy is likely due to the poor polynomial fit of the MPT curve
(Fig. \ref{fig:real_eg_3_MSD_Gp_Gpp} (a)), which is dominated by the large lag time values. Potential improvements include truncating the MPT or downweighting the fit on large lag time points due to large uncertainty in the estimates. These results indicate that the low-frequency moduli estimates from microrheology estimation are sensitive to the shape of the MSD curve at larger lag time points due to the limited information from microscopy videos, and should be interpreted with caution. For MPT, this sensitivity is further affected by the choice of the smoothing procedure,  as is evident by the substantial difference between the moduli obtained from spline smoothing and those obtained from polynomial approximation. In comparison, MF-AIUQ can be applied to GSER context by providing smooth MSD estimates, thereby removing the need for the additional smoothing step required in MPT.

\section{DISCUSSION}

This works presents MF-AIUQ, a model-free estimation of dynamical properties, including the mean squared displacement curve over the entire domain, without the need of segmenting and linking individual particle trajectories. MF-AIUQ uses the relationship between the ISF and MSD derived from the cumulant expansion \cite{koppel1972analysis, nijboer1966time}, enabling direct estimation of the MSD through marginal maximum likelihood. To reduce the computational cost of likelihood evaluation,  Fourier-transformed values at both the spatial and temporal domains are subsampled, and the generalized Schur algorithm \cite{gohberg1972inversion,ammar1988superfast, ling2019superfast} is applied for efficient covariance matrix inversion and log determinant calculation. The resulting approximately log equally spaced wave vectors and lag times retain the main information needed for likelihood-based MSD estimation. The estimated MSD  can be further applied to estimate viscoelastic moduli through the GSER. Furthermore, MF-AIUQ also provides uncertainty quantification for the MSD estimates by accounting for two sources of uncertainty, including Fourier discretization and parameter estimation error. 

The simulation studies and experimental analysis confirm that MF-AIUQ provides reliable MSD estimates without requiring particle tracking or specification of a dynamic model. The three experimental systems differ in particle size, imaging contrast, and particle dynamics, and together they provide an evaluation of MF-AIUQ under diverse  conditions. Furthermore, the computational cost of MF-AIUQ does not increase compared to MD-AIUQ, due to the proposed subsampling strategy that offsets the cost for estimating the  MSD curves over the entire lag time range.

Nevertheless, several limitations remain. First, the current framework does not accommodate drift, which may arise in some experimental settings. Second, although storage and loss moduli can be estimated by the MF-AIUQ approach, the estimation of a small frequency range contains large uncertainty due to the limited information of image pairs with a large lag time. Third, the current uncertainty analysis does not incorporate interpolation uncertainty from the GPR step, and the total uncertainty is currently obtained through direct summation of contributions from Fourier discretization and parameter estimation. Developing a more principled approach to combining these sources of additional uncertainty remains an open problem. 

\section{Acknowledgement}
This work is supported by the National Science Foundation (NSF) award No. OAC-2411043 and No. OAC-2411044, with partial support from NSF BioPACIFIC MIP under Award No. 2445868, and NSF Materials Research Science and Engineering Center (MRSEC) under Award No. DMR-2308708 (IRG-1).

\bibliography{References_2024}

\appendix

\section{\label{app:simulation}Simulation for Isotropic Processes}

The six simulated videos in Section \ref{sec:simulation} are generated based on the simulation introduced in Appendix A and also in Ref. \cite{gu2024ab}. For $m = 1,\dots, M$ and $k = 1, \dots, n$, let $\mathbf{x}^{(m)}(t_k) = \bigl(x^{(m)}_1(t_k),\, x^{(m)}_2(t_k)\bigr)$ denote the 2D location of particle $m$ at time $t_k$, where $x^{(m)}_1$ and $x^{(m)}_2$ represent the horizontal and vertical coordinates, respectively. Let $\mathbf{x}^{(m)}(0)=\bigl(x^{(m)}_1(0),\, x^{(m)}_2(0)\bigr)$ denote the initial location of particle $m$. All simulated processes are isotropic.

Let $N_1$ and $N_2$ denote the horizontal and vertical frame sizes, respectively. The initial positions are sampled separately from the central region of the image:
\begin{align*}
    & x^{(m)}_1(0)\sim \mathrm{Unif}\left(\frac{N_1}{8},\frac{7N_1}{8}\right), \\
    & x^{(m)}_2(0)\sim \mathrm{Unif}\left(\frac{N_2}{8},\frac{7N_2}{8}\right),
\end{align*}
for $m=1,\dots,M$.

Given the initial locations, the two spatial coordinates are generated independently and then combined to form the 2D trajectories.

For Brownian motion, the trajectory is generated by adding independent Gaussian increments at each time point:
\begin{align*}
 & \mathbf{x}^{(m)}(t_1) = \mathbf{x}^{(m)}(0), \\
 & \mathbf{x}^{(m)}(t_k)=\mathbf{x}^{(m)}(t_{k-1})+ \frac{\sigma_{BM}^2}{2} \boldsymbol{\xi}^{(m)}(t_k),
\end{align*}
for $k = 2,\dots,n$, where $\boldsymbol{\xi}^{(m)}(t_k)\sim \mathcal{MN}\bigl(\mathbf{0},\mathbf I_2\bigr)$, and $\mathbf I_2$ is the $2\times 2$ identity matrix. The corresponding true MSD at lag time point $\Delta t_k$ is denoted by $\theta_{true,\Delta t_k}$, for $k = 1,\dots,n-1$, it follows
\[
\theta_{true,\Delta t_k}=\sigma_{BM}^2 \Delta t_k.
\]

For the Ornstein-Uhlenbeck process, the particle position at the first time frame is obtained by adding Gaussian noise to the initial position,
\[
\mathbf{x}^{(m)}(t_1)=\mathbf{x}^{(m)}(0)+\frac{\sigma_{OU}^2}{4} \bm\xi^{(m)}(t_1),
\]
where $\bm\xi^{(m)}(t_1)\sim \mathcal{MN}\bigl(\mathbf{0},\mathbf I_2\bigr)$. Then for $k = 2,\dots,n$,
\begin{align*}
\mathbf{x}^{(m)}(t_k)
& = \mathbf{x}^{(m)}(0)+
\rho\bigl(\mathbf{x}^{(m)}(t_{k-1})-\mathbf{x}^{(m)}(0)\bigr) \\
& \quad + \frac{\sigma_{OU}^2(1-\rho^2)}{4}\bm\xi^{(m)}(t_k),
\end{align*}
The corresponding true MSD at lag $\Delta t_k$ is
\[
\theta_{true,\Delta t_k}=
\sigma_{OU}^2\bigl(1-\rho^{\Delta t_k}\bigr).
\]

For the fractional Brownian motion process, let $\mathbf{x}^{(m)}(t_1)=\mathbf{x}^{(m)}(0)$. For each coordinate direction $l=1,2$, we have
\[
\Delta x^{(m)}_l(t_k)=x^{(m)}_l(t_k)-x^{(m)}_l(t_{k-1}).
\]
The increment vector for each coordinate direction,
\[
\bigl(\Delta x^{(m)}_l(t_2),\dots,\Delta x^{(m)}_l(t_n)\bigr)^T,
\]
is generated from a multivariate normal distribution with mean zero and covariance matrix $\frac{\sigma_{FBM}^2}{2}\Sigma_H$, where the $(r,s)$th entry of $\Sigma_H$ is given by
\begin{align*}
\Bigl[\Sigma_H\Bigr]_{r,s}
&=
\frac{1}{2}|t_r-t_s+\Delta t_{min}|^{2H}
+\frac{1}{2}|t_r-t_s-\Delta t_{min}|^{2H} \\
&\quad
-|t_r-t_s|^{2H},
\end{align*}
for $r,s=2,\dots,n$. Thus, the increments are correlated across time, with dependence determined by the Hurst parameter $H=\alpha/2$. The corresponding true MSD at lag $\Delta t_k$ is
\[
\theta_{true,\Delta t_k}=
\sigma_{FBM}^2(\Delta t_k)^{2H}.
\]

For the mixture of OU and FBM process, we first add a Gaussian noise to the initial position:
\[
\mathbf{x}^{(m)}(t_1)=\mathbf{x}^{(m)}(0)+ \frac{\sigma_{OU}^2}{4}\bm{\xi}^{(m)}(t_1).
\]
The OU and FBM components of particle $m$ follow
\[
\mathbf{x}^{OU,(m)}(t_1)=\mathbf{x}^{FBM,(m)}(t_1)=\mathbf{x}^{(m)}(t_1).
\]
For $k=2,\dots,n$, the OU component is generated by
\begin{align*}
   \mathbf{x}^{OU,(m)}(t_k)
   & = \mathbf{x}^{(m)}(t_1)+
\rho\bigl(\mathbf{x}^{OU,(m)}(t_{k-1})-\mathbf{x}^{(m)}(t_1)\bigr) \\
   & \quad + \frac{\sigma_{OU}^2(1-\rho^2)}{4} \bm{\xi}^{OU,(m)}(t_k), 
\end{align*}
The FBM component is generated as in the FBM model above. For $k=1,\dots,n$, the trajectory is given by
\[
\mathbf{x}^{(m)}(t_k)=\mathbf{x}^{OU,(m)}(t_k)+\mathbf{x}^{FBM,(m)}(t_k)-\mathbf{x}^{(m)}(t_1).
\]
The corresponding true MSD at lag $\Delta t_k$ is
\[
\theta_{true,\Delta t_k}=
\sigma_{OU}^2\bigl(1-\rho^{\Delta t_k}\bigr)+
\sigma_{FBM}^2(\Delta t_k)^{2H}.
\]

After generating the particle trajectories, each particle is represented on the image by an intensity function centered at its simulated location, where the intensity decays exponentially with the squared distance from the particle center. For particle $m$ at time $t_k$, its contribution to the intensity at location $\mathbf{x}_{j_1,j_2}$, with $j_1 = 1,\dots, N_1$ and $j_2 = 1,\dots, N_2$, is defined by
\begin{align*}
Y^{(m)}_{j_1,j_2}(t_k)
& = Y_{max}\exp\left\{
-\frac{\|\mathbf{x}_{j_1,j_2}-\mathbf{x}^{(m)}(t_k)\|^2}{2\sigma_p^2}
\right\} \\
& \quad {1}_{\!
\|\mathbf{x}_{j_1,j_2}-\mathbf{x}^{(m)}(t_k)\|\le 3\sigma_p},
\end{align*}
where $Y_{max}$ denotes the peak intensity, $\sigma_p$ is a parameter controlling the radius of the spherical particle. The intensity at location $\mathbf{x}_{j_1,j_2}$ and time $t_k$ is then obtained by adding the contributions from all particles and the Gaussian background noise:
\[
Y_{j_1,j_2}(t_k)
=
\sum_{m=1}^M Y^{(m)}_{j_1,j_2}(t_k)+\epsilon_{j_1,j_2}(t_k),
\]
where $\epsilon_{j_1,j_2}(t_k)\sim N(0,\frac{B}{2})$ denotes the background Gaussian white noise with variance $B/2$.

\section{\label{app:MSD_uncertainty}Uncertainty Quantification}

\begin{figure*}[t]
\refstepcounter{algocf}
\noindent\rule{\textwidth}{1pt}\\
\textbf{Algorithm \thealgocf:} MF-AIUQ for MSD estimation and uncertainty quantification\label{alg:MF_AIUQ}\\
\rule{\textwidth}{1pt}
\begin{minipage}[t]{0.48\textwidth}
\footnotesize
\begin{algorithmic}[1]

\REQUIRE Microscopy file of size $N_1\times N_2$, recorded over $n$ time frames
\ENSURE MSD estimates $\{\theta_{est, \Delta t_k}\}_{k=1}^{n-1}$; optional $95\%$ confidence intervals $\{[\theta^{L}_{est, \Delta t_k}, \theta^{U}_{est, \Delta t_k}]\}_{k=1}^{n-1}$
\vspace{3pt}
\STATE \textbf{Pre-processing Phase}
\vspace{3pt}

\STATE Normalize the range of intensity to be $[0,1]$.
\STATE Transform intensities at each time frame to reciprocal space via 2D FFT.

\STATE \textit{Spatial frequency subsampling:}
\STATE \quad (i) $J_0 \leftarrow \min\left\{k:\sum_{j=1}^{k}A_j\big/\sum_{j=1}^{J}A_j\ge 1-\varepsilon_1\right\}$.
\STATE \quad For $j =J_0+1, \dots, J$:
\STATE \quad \quad If $A_j \ge \varepsilon_2$, set $J_0 \leftarrow j$.
\STATE \quad (ii) Subsample frequencies with index spacing defined in Eq.~\eqref{equ:delta_log_q}, indexed by $\mathcal{J}_s$.

\STATE \textit{Lag time subsampling:}
\STATE \quad Subsample lag times with index spacing defined in Eq.~\eqref{equ:delta_log_dt}, indexed by $\mathcal{T}_s$.

\vspace{3pt}
\STATE \textbf{Optimization and Estimation Phase}
\vspace{3pt}

\STATE \textit{Stage 1:}

\end{algorithmic}
\end{minipage}
\hfill
\begin{minipage}[t]{0.48\textwidth}
\footnotesize
\begin{algorithmic}[1]
\setcounter{ALC@line}{12}

\STATE \quad (i) Initialize at 
$\tilde{\bm\theta}^{(0)}_{\mathcal{T}_s} \leftarrow$ Eq.~\eqref{equ:ini_tilde_theta},
$\quad \tilde{B}^{(0)} \leftarrow$ Eq. (\ref{equ:B_ini}).
\STATE \quad (ii) In each iteration, predict MSD over the lag time range with spacing $c\Delta t_{min}$ indexed by $\mathcal{T}_c$ using GPR.
\STATE \quad (iii) Obtain optimized values:
\STATE \quad \quad
$(\tilde{\bm \theta}^{c}_{est,\mathcal{T}_s},\, \tilde{B}^{c}_{est})
\leftarrow
\argmax_{\tilde{\bm \theta}_{\mathcal{T}_s},\, \tilde{B}}
\mathcal{L}_{\mathcal T_c,\,\mathcal J_s}
\!\left(\tilde{\bm \theta}_{\mathcal{T}_s}, \mathbf{A}_{est,\mathcal J_s}, \tilde{B}\right).$

\STATE \textit{Stage 2:}
\STATE \quad (i) Initialize at $\{\tilde{\bm \theta}^{c}_{est,\mathcal{T}_s},\, \tilde{B}^{c}_{est}\}$.
\STATE \quad (ii) In each iteration, predict MSD over the entire lag time range indexed by $\mathcal{T}$ using GPR.
\STATE \quad (iii) Obtain optimized values:
\STATE \quad \quad
$(\tilde{\bm \theta}_{est,\mathcal{T}_s},\, \tilde{B}_{est})
\leftarrow
\argmax_{\tilde{\bm \theta}_{\mathcal{T}_s},\, \tilde{B}}
\mathcal{L}_{\mathcal{T},\,\mathcal{J}_s}
\!\left(\tilde{\bm \theta}_{\mathcal{T}_s}, \mathbf{A}_{est, \mathcal{J}_s}, \tilde{B}\right).$

\STATE Predict MSD over $\mathcal{T}$ via GPR.

\vspace{3pt}
\STATE \textbf{Optional: Uncertainty Quantification Phase}
\vspace{3pt}

\STATE Obtain the $95\%$ confidence intervals for $\tilde{\bm \theta}_{\mathcal{T}_s}$ via Eqs.~\eqref{equ:Toeplitz_oprator}, \eqref{equ:obs_Fisher_compute}, \eqref{equ:asymptotic_sd}, and \eqref{equ:95CI}.

\STATE $[\bm \theta^L_{est},\, \bm \theta^U_{est}] \leftarrow$ GPR and exponential transformation.

\RETURN $\{\theta_{est, \Delta t_k}\}_{k=1}^{n-1}$ and $\{[\theta^{L}_{est, \Delta t_k}, \theta^{U}_{est, \Delta t_k}]\}_{k=1}^{n-1}$.

\end{algorithmic}
\end{minipage}\\
\rule{\textwidth}{1pt}
\end{figure*}

Denote $\bm{\psi}_{est} = [\tilde{\bm{\theta}}_{est,\, \mathcal{T}_s}^T,\, \tilde{B}_{est}]^T \in \mathbb{R}^{\tilde{n}_s+1}$ as the MMLE of the logarithmic MSD curves and variance of the noise in intensity obtained from the two-stage optimization. To quantify uncertainty in the MF-AIUQ estimates, we account for two sources of uncertainty. The first is associated with the discrete Fourier transform. While the wave vector is continuous in theory, only discrete values are considered in the analysis. To evaluate the effect of this discretization, we examine how the parameter estimates change with small perturbations of the wave vector magnitudes. Specifically, let $\bm{\psi}_{est}^{q-q_{min}}$ denote the MMLE obtained after replacing each wave vector magnitude $q_j$ by $q_j - q_{min}$, and let $\bm{\psi}_{est}^{q+q_{min}}$ denote the MMLE obtained after replacing each $q_j$ by $q_j + q_{min}$, where $q_{min}$ is the minimum wave vector magnitude. The uncertainty of discretization is given by elementwise lower and upper bounds formed from $\bm{\psi}_{est}^{q-q_{min}}$ and $\bm{\psi}_{est}^{q+q_{min}}$, and then further adjusted elementwise, so that each resulting interval contains the corresponding MMLE $\bm{\psi}_{est}$. Specifically, for each component $k$, we define
\begin{align*}
\psi_{est,k}^{q,L}
&=
\min\{\psi_{est,k}^{q-q_{\min}},\,\psi_{est,k}^{q+q_{\min}},\,\psi_{est,k}\},\\
\psi_{est,k}^{q,U}
&=
\max\{\psi_{est,k}^{q-q_{\min}},\,\psi_{est,k}^{q+q_{\min}},\,\psi_{est,k}\}.
\end{align*}

The second source of uncertainty comes from parameter estimation based on a finite sample. We approximate the sampling variability of the maximum marginal likelihood estimator using asymptotic theory, since the number of pixels and time points are typically large. Under regularity conditions, $\bm{\psi}_{est}$ is asymptotically normal \cite{mardia1984maximum}
\[
\sqrt{M} (\bm{\psi}_{est} - \bm{\psi}_0)
\xrightarrow{d} \mathcal{N} \left(0,\, \mathcal{I}^{-1}_{\bm{\psi}_0} \right),
\]
where $\bm{\psi}_0 \in \mathbb{R}^{\tilde n_s+1}$ denotes the unknown true parameter vector, and $\mathcal{I}_{\bm{\psi}_0}$ is the $(\tilde n_s+1)\times (\tilde n_s+1)$ Fisher information matrix. In our setting, $M$ is used as the effective sample size, since the scattering signal is determined by the displacement of the $M$ particles, whereas the number of Fourier pixels can be substantially larger. The corresponding rescaling in the practical construction of the observed Fisher information is described below.

The Fisher information matrix $\mathcal{I}_{\bm{\psi}_0}$ quantifies the amount of information that the observations carry about $\bm{\psi}$, where $\bm{\psi} = [\tilde{\bm{\theta}}_{\mathcal{T}_s}, \tilde{B}]^T$, with the $(r,s)$th entry defined as
\begin{equation}
\Bigl[\mathcal{I}_{\bm{\psi}_0}\Bigr]_{r,s} = -\mathbb{E}\left[\frac{\partial^2 \ell_{\mathcal{T}, \mathcal{J}_0}(\bm{\psi}, \mathbf{A}_{\mathcal{J}_0})}{\partial \psi_r \partial \psi_s}\Big|_{\bm{\psi} = \bm{\psi}_0}\right],
\label{equ:Fisher_mat}
\end{equation}
where $r,s = 1,\dots, \tilde{n}_s+1$, and $\psi_r$ and $\psi_s$ denote the $r$th and $s$th elements of $\bm{\psi}$, respectively. Here, $\ell_{\mathcal{T}, \mathcal{J}_0}(\bm{\psi}, \mathbf{A}_{\mathcal{J}_0})$ denotes the log-likelihood over all lag times indexed by $\mathcal{T}$ and spatial frequencies indexed by $\mathcal{J}_0$, where $\mathcal{J}_0 = \{1, 2, \dots, J_0 \}$. The log-likelihood $\ell_{\mathcal{T}, \mathcal{J}_0}(\bm{\psi}, \mathbf{A}_{\mathcal{J}_0})$ is
\begin{align*}
    \ell_{\mathcal{T}, \mathcal{J}_0}(\bm{\psi}, \mathbf{A}_{\mathcal{J}_0})
    &=  -n\sum_{j=1}^{J_0}N_{S_j}\log 2\pi -\sum_{j=1}^{J_0} N_{S_j}\log |\bm{\Sigma}_{\bm{\psi},\, j}| \\
    & \quad - \frac{1}{2} \sum_{j=1}^{J_0}\sum_{\mathbf{j}' \in \mathcal{S}_j}\Bigl[\hat{\mathbf{y}}_{re,\mathbf{j}'}^T \bm{\Sigma}^{-1}_{\bm{\psi},\, j} \hat{\mathbf{y}}_{re,\mathbf{j}'} \\
    & \qquad \qquad + \hat{\mathbf{y}}_{im,\mathbf{j}'}^T \bm{\Sigma}^{-1}_{\bm{\psi}, \, j} \hat{\mathbf{y}}_{im,\mathbf{j}'} \Bigr].
\end{align*}
The covariance matrix for the real or imaginary part of the Fourier transformed intensity on the $j$th ring, denoted by $\bm{\Sigma}_{\bm{\psi},\, j}$, is defined in Eq.~(\ref{equ:covariance_mat}), with the $(r,s)$th entry given by
\[
\Bigl[\Sigma_{\bm{\psi},\, j}\Bigr]_{rs}
= \frac{A_{j}}{4} \exp\!\left(
-\frac{q_j^2\, \theta_{\Delta t_{|r-s|}}}{4}
\right) + \frac{B}{4} {1}_{|r-s| = 0},
\]
where $r,s = 1,\dots, n$, $\Delta t_{|r-s|}$ is the lag time corresponding to index difference $|r-s|$, $\theta_{\Delta t_{|r-s|}}$ is the MSD at that lag time, and $\theta_{\Delta t_0} = 0$ since the MSD at zero lag time is zero. The $(r,s)$th entry of the Fisher information then takes the form
\[
\Bigl[\mathcal{I}_{\bm{\psi}_0}\Bigr]_{rs}
=
\sum_{j=1}^{J_0} N_{S_j}\,
\operatorname{tr} \Biggl(
\bm{\Sigma}^{-1}_{\bm{\psi},\, j}
\frac{\partial \bm{\Sigma}_{\bm{\psi},\, j}}{\partial \psi_r}
\bm{\Sigma}^{-1}_{\bm{\psi},\, j}
\frac{\partial \bm{\Sigma}_{\bm{\psi},\, j}}{\partial \psi_s}
\Biggr)\Bigg|_{\bm{\psi}=\bm{\psi}_0}.
\]

In practice, $\mathcal{I}_{\bm{\psi}_0}$ depends on the unknown $\bm{\psi}_0$, so it cannot be computed directly. We therefore use the observed Fisher information matrix $\mathcal{I}^{obs}_{\bm{\psi}_{est}}$, obtained by evaluating the Fisher information expression at the MMLE $\bm{\psi}_{est}$, with the $(r,s)$th entry given by
\[
\Bigl[\mathcal{I}^{obs}_{\bm{\psi}_{est}}\Bigr]_{rs} = \sum_{j=1}^{J_0} N_{S_j}
\operatorname{tr} \Biggl(
\bm{\Sigma}^{-1}_{\bm{\psi}, j}
\frac{\partial \bm{\Sigma}_{\bm{\psi}, j}}{\partial \psi_r}
\bm{\Sigma}^{-1}_{\bm{\psi}, j}
\frac{\partial \bm{\Sigma}_{\bm{\psi}, j}}{\partial \psi_s}
\Biggr)\Bigg|_{\bm{\psi}=\bm{\psi}_{est}}.
\]
Here $\bm{\Sigma}_{\bm{\psi}, j}|_{\bm\psi = \bm\psi_{est}}$ uses the GPR interpolated MSD estimates $\bm{\theta}_{est}$ over the entire lag time domain. Since $\sum_{j=1}^{J_0} N_{S_j}$ is large, $\mathcal{I}^{obs}_{\bm{\psi}_{est}}$ is used to approximate $\mathcal{I}_{\bm{\psi}_0}$, so the asymptotic covariance of $\bm{\psi}_{est}$ is approximated by $(\mathcal{I}^{obs}_{\bm{\psi}_{est}})^{-1}$.

To obtain $\mathcal{I}^{obs}_{\bm{\psi}_{est}}$, we first calculate the derivative of $\bm{\Sigma}_{\bm{\psi}, j}$ with respect to each parameter in $\bm{\psi}$, evaluated at $\bm{\psi}_{est}$. For $k = 1,\dots, \tilde{n}_s$, the $(r,s)$th entry of this derivative is
\begin{align*}
\Bigl[\frac{\partial \bm{\Sigma}_{\bm{\psi},j}}{\partial \tilde{\theta}_{\Delta \tilde{t}_k}}\Big|_{\bm{\psi}=\bm{\psi}_{est}}\Bigr]_{rs}
&= \exp\!\left(-\frac{q_j^2 \theta_{est,\Delta t_{|r-s|}}}{4}\right) \\
&\quad \times \left[\frac{\partial \bm{\theta}^{aug}_{est}}{\partial \tilde{\theta}_{est,\Delta \tilde{t}_k}}\right]_{|r-s|} \frac{A_{est,j}}{4}\left(-\frac{q_j^2}{4}\right),
\end{align*}
where $\bm{\theta}^{aug}_{est} = (0, \theta_{est,\Delta t_1}, \dots, \theta_{est,\Delta t_{n-1}})^T$ denotes the augmented MSD vector over the full lag time domain, obtained by adding $0$ to the GPR interpolated MSD estimates $\bm{\theta}_{est}$, since the MSD at $\Delta t_0$ is $0$. For notation simplicity, $\partial \bm{\theta}^{aug}_{est}/\partial \tilde{\theta}_{est,\Delta \tilde{t}_k}$ denotes the derivative vector evaluated at $\bm{\psi}=\bm{\psi}_{est}$, and $\Bigl[\partial \bm{\theta}^{aug}_{est}/\partial \tilde{\theta}_{est,\Delta \tilde{t}_k}\Bigr]_{|r-s|}$ represents its element corresponding to index difference $|r-s|$, which is computed numerically.

The $(r,s)$th entry of the derivative of $\bm{\Sigma}_{\bm{\psi},j}$ with respect to $\tilde{B}$, evaluated at $\bm{\psi}_{est}$, is
\begin{align*}
   \Bigl[ \frac{\partial\,\bm{\Sigma}_{\bm{\psi},j}}{\partial\, \tilde{B}} \Big|_{\bm{\psi}=\bm{\psi}_{est}} \Bigr]_{rs} 
   & = -\frac{B_{est}}{4}\exp\!\Bigl(-\frac{q_j^2\theta_{est,\Delta t_{|r-s|}}}{4}\Bigr) \\
   & \quad + \frac{B_{est}}{4}{1}_{|r-s|=0}.
\end{align*}

The observed Fisher information is assembled across the rings using a Toeplitz trace operator $\mathbf{T}_z(\cdot,\cdot)$ from the SuperGauss package in R \cite{Ling2022SuperGuasspackage}. For ring $j$, the contribution to $\mathcal{I}^{obs}_{\bm{\psi}_{est}}$ from a single pixel is an $(\tilde n_s+1)\times(\tilde n_s + 1)$ matrix
\begin{equation}
  \mathcal{I}^{obs}_{\bm{\psi}_{est},\,j}
=
\biggl[
\mathbf{T}_z\Bigl(\frac{\partial \bm{\gamma}_{\bm{\psi},j}}{\partial \psi_r}\Big|_{\bm{\psi}=\bm{\psi}_{est}}, \frac{\partial \bm{\gamma}_{\bm{\psi},j}}{\partial \psi_s}\Big|_{\bm{\psi}=\bm{\psi}_{est}} \Bigr)
\biggr]_{r,s=1}^{\tilde n_s+1},
\label{equ:Toeplitz_oprator}
\end{equation}
where $\bm{\gamma}_{\bm{\psi},j}$ is the autocovariance vector for the real or imaginary part of the Fourier transform intensity on the $j$th ring, given by the first row of $\bm{\Sigma}_{\bm{\psi},j}$. Since the number of Fourier pixels $\sum_{j=1}^{J_0}N_{S_j}$ is typically much larger than the number of particles $M$, using this quantity directly would overestimate the information available for the asymptotic approximation. We therefore rescale the average observed information from the Fourier pixels to the effective sample size $M$. The total observed Fisher information is then
\begin{equation}
    \mathcal{I}^{obs}_{\bm{\psi}_{est}} = \frac{M}{\sum_{j=1}^{J_0} N_{S_j}}
\sum_{j=1}^{J_0} 
N_{S_j}\,\mathcal{I}^{obs}_{\bm{\psi}_{est},\,j}.
\label{equ:obs_Fisher_compute}
\end{equation}

The asymptotic standard deviations of the estimated parameters, denoted by $\bm{\sigma}_{\bm{\psi}_{est}}$, can then be estimated based on the observed Fisher information, with the $k$th element given by
\begin{equation}
    \sigma_{\psi_{est, k}} = \sqrt{\bigl[(\mathcal{I}^{obs}_{\bm{\psi}_{est}} )^{-1}\bigr]_{kk}}, \quad k = 1,\dots, \tilde n_s+1.
\label{equ:asymptotic_sd}
\end{equation}

The asymptotic standard deviations are then used to construct the $95\%$ confidence intervals. Let $\bm{\psi}_{est}^{L}$ and $\bm{\psi}_{est}^{U}$ denote the resulting lower and upper bounds for $\bm{\psi}_{est}$, respectively. The $k$th elements are given by
\begin{align}
    \psi_{est,k}^{L}
    &=
    \psi_{est,k}^{q,L}-\zeta_{0.975}\sigma_{\psi_{est,k}}, \nonumber\\
    \psi_{est,k}^{U}
    &=  \psi_{est,k}^{q,U}+\zeta_{0.975}\sigma_{\psi_{est,k}},
\label{equ:95CI}
\end{align}
where $\zeta_{0.975}$ is the 0.975 quantile of the standard normal distribution.

The upper and lower bounds of the $95\%$ confidence intervals for the MSD estimates over the full lag time range are obtained by GPR interpolation and exponential transformation.
The resulting upper and lower bounds are denoted as $\bm{\theta}_{est}^{U}$ and $\bm{\theta}_{est}^{L}$, respectively.

\section{\label{app:MSD_Gp_Gpp}Storage and Loss Moduli Estimation}
\begin{algorithm}[th]
\caption{Moduli estimation by GSER}
\label{alg:gser_moduli}
\begin{algorithmic}[1]
\REQUIRE $\{[\theta^L_{est,\,\Delta t_k},\, \theta^U_{est,\,\Delta t_k}]\}_{k=1}^{n-1}$, $r$, $T_a$, $k_B$
\ENSURE $\{G'_{est,\, \omega_k}, G''_{est,\, \omega_k}\}_{k=1}^{n-1}$

  \STATE Draw 1000 samples of log MSD curves via Eq. \eqref{equ:log_MSD_distribution}.
  \STATE Transform each sampled log MSD curve to the original scale.

  \STATE For $l = 1, \dots, 1000$:
  \STATE \quad For $k = 2, \dots, n-2$: \\
  \qquad $\alpha^{(l)}_{\Delta t_k} \leftarrow$ Eq. \eqref{equ:alpha_MF_AIUQ_1} using the $l$th sampled MSD curve.
  \STATE \quad For $k = 1\quad \text{or}\quad n-1$: \\
  \qquad $\alpha^{(l)}_{\Delta t_k} \leftarrow$ Eqs. \eqref{equ:alpha_MF_AIUQ_2} and \eqref{equ:alpha_MF_AIUQ_3} using the $l$th sampled MSD curve.
  \vspace{0.3em}

  \STATE \quad For $k =1,\dots, n-1$: \\
  \quad \quad (i) Compute frequency $\omega_k \leftarrow 1 / \Delta t_k$. \\
  \quad \quad (ii) Compute the magnitude of the complex modulus $|G^{*(l)}_{\omega_k}|$ by Eq. \eqref{equ:G_complex}. \\
  \quad \quad (iii) Compute storage modulus and loss modulus $G'^{(l)}_{\omega_k}$ and $G''^{(l)}_{\omega_k}$ by Eqs. \eqref{equ:Gp} and \eqref{equ:Gpp}.
  \vspace{0.5em}

  \STATE For $k =1,\dots,n-1$: \\
  \quad $G'_{est,\,\omega_k}
  \leftarrow
  \operatorname{med}\{G'^{(l)}_{\omega_k}: l=1,\dots,1000\}$ \\
  \quad $G''_{est,\,\omega_k}
  \leftarrow
  \operatorname{med}\{G''^{(l)}_{\omega_k}: l=1,\dots,1000\}$

\RETURN $\{G'_{est,\, \omega_k}, G''_{est,\, \omega_k}\}_{k=1}^{n-1}$
\end{algorithmic}
\end{algorithm}

Given that the transformation from the MSD to the storage and loss moduli through the GSER is nonlinear, the moduli calculated directly from the point estimate of the MSD do not necessarily correspond to the predictive means of the distributions of $G'_{\omega}$ and $G''_{\omega}$. Therefore, instead of transforming only the estimated MSD curve, we propagate uncertainty in the MSD by Monte Carlo sampling.

Assume that log MSD $\tilde{\bm \theta} = [\tilde{\theta}_{\Delta t_1}, \dots, \tilde{\theta}_{\Delta t_{n-1}}]^T$ follows
\begin{equation}
   \tilde{\bm \theta} \sim \mathcal{MN}(\bm \mu_{\tilde{\bm \theta}}, \mathbf{\Sigma}_{\tilde{\bm \theta}}).
   \label{equ:log_MSD_distribution}
\end{equation}
 
Let $\tilde{\bm \theta}_{est}^U = \log \bm \theta_{est}^U$ and $\tilde{\bm \theta}_{est}^L = \log \bm \theta_{est}^L$, the mean vector is taken as 
\[
\bm \mu_{\tilde{\bm \theta}} = \frac{\tilde{\bm \theta}_{est}^L +
\tilde{\bm \theta}_{est}^U}{2}, 
\]
where $\bm \mu_{\tilde{\bm \theta}}$ corresponds to the median of the $95\%$ confidence interval of MSD estimates.

The pointwise standard deviation of log MSD is estimated from the $95\%$ uncertainty interval:
\[
\sigma_{\tilde{\bm \theta}, k} =
\frac{\tilde{\bm \theta}_{est,k}^U - \tilde{\bm \theta}_{est,k}^L}{2 \zeta_{0.975}},
\]
where $k=1,\dots, n-1$, and $\zeta_{0.975}$ represents the 0.975 quantile of standard normal distribution. The $(r,s)$th entry of the covariance matrix is given by 
\[
\left[\Sigma_{\tilde{\bm \theta}} \right]_{r,s} = \sigma_{\tilde{\bm \theta}, r} \sigma_{\tilde{\bm \theta}, s} K(\log \Delta t_r,\, \log \Delta t_s),
\]
where $K(\cdot,\cdot)$ is the Mat\'ern correlation kernel with the roughness parameter being 2.5 \cite{rasmussen2006gaussian}.

We then draw 1000 samples of log MSD from Eq. (\ref{equ:log_MSD_distribution}). Each sampled curve is transformed back to the original scale and used to calculate the storage and loss moduli. Denote $\mathbf G'^{(j)} = [G'^{(j)}_{\omega_1},\dots,G'^{(j)}_{\omega_{n-1}}]^T$ and $\mathbf G''^{(j)} = [G''^{(j)}_{\omega_1},\dots,G''^{(j)}_{\omega_{n-1}}]^T$, $j=1,\dots, 1000$, as the storage and loss moduli obtained from the $j$th sampled log MSD curve. We use the pointwise sample medians as the estimator, with the $k$th estimate given by:
\begin{align}
    & G'_{est,\,\omega_k} = \operatorname{med}\left\{G'^{(j)}_{\omega_k}:\, j=1,\dots,1000 \right\}, \nonumber \\
    & G''_{est,\,\omega_k} = \operatorname{med}\left\{G''^{(j)}_{\omega_k}:\, j=1,\dots,1000 \right\},
    \label{equ:Gp_Gpp_estimate}
\end{align}
where $k = 1,\dots, n-1$, and $\operatorname{med}\{\cdot\}$ represents the median operator.

\section{\label{app:computational_cost} Comparison of the Computational Cost}

We compare the computation cost of the three DDM-based methods, MF-AIUQ, MD-AIUQ, and DDM-UQ, using four examples, including simulated BM with $\sigma_{BM}^2 = 0.02$, the PVA solution with $1\, \mu m$ particles, tetraPEG hydrogel at 13 min, and the snail mucin experiment. All computations are performed on a macOS 15.5 system with an Apple M1 chip, eight CPU cores, and 16 GB of RAM.

\begin{figure}[H]
\vspace{.05in}
    \centering
    \includegraphics[scale=0.48]{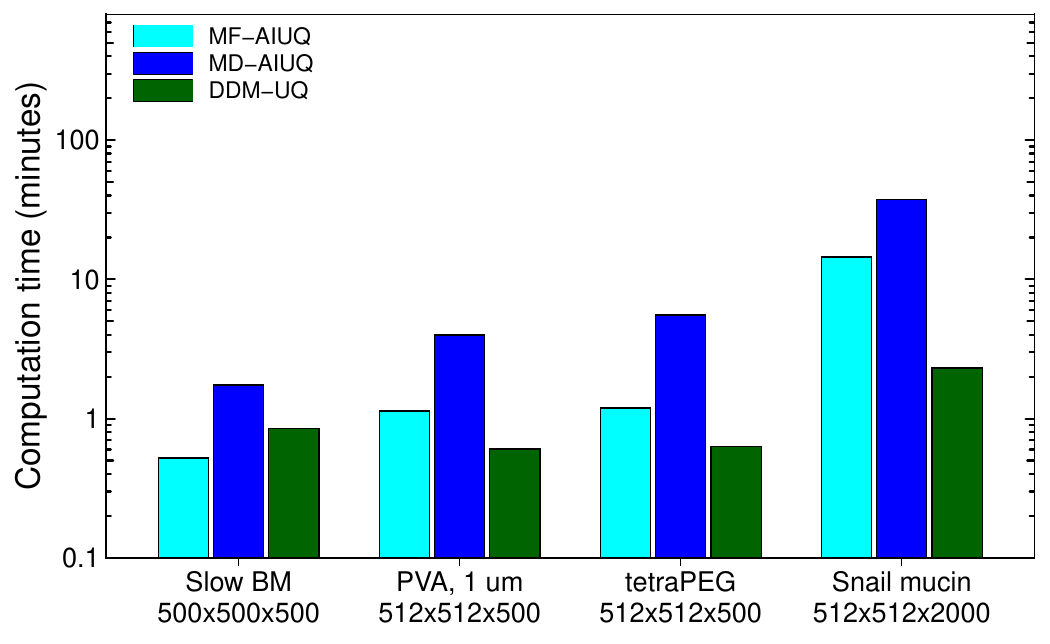}
    \caption{Computation time for MF-AIUQ, MD-AIUQ, and DDM-UQ. The examples include BM with $\sigma_{BM}^2 = 0.02$, $4\%$ PVA solution with $1\, \mu m$ particles, tetraPEG hydrogel at 13 min, and snail mucin. The image stack size $N_1\times N_2 \times n$ is labeled below each data set.}
    \label{fig:computation_time}
\end{figure}

For the simulation example, MF-AIUQ has the shortest computation time, while for the other three cases, DDM-UQ is the most efficient, followed by MF-AIUQ (Fig. \ref{fig:computation_time}). The low computational cost of DDM-UQ stems from computing the Fourier transform of image differences only at a small subset of lag times, which has been shown to reduce computational cost by more than 100 times compared with conventional DDM for image sequences of size $512 \times 512$ over 1000 time frames \cite{gu2021uncertainty}. However, DDM-UQ provides valid MSD estimates only over a limited lag time range in many scenarios. In MF-AIUQ, we reduce the computational cost by subsampling both spatial frequencies and lag time points, making it more efficient than MD-AIUQ. Although MF-AIUQ remains more computationally expensive than DDM-UQ in three of these examples, it provides smooth MSD estimates over the full lag time range, which are crucial for some studies.

\end{document}